\documentclass[useAMS,usenatbib]{mnras}

\usepackage{graphicx}
\usepackage{txfonts}

\title[Regularity of the young stellar population]
      {Spatial regularity of the young stellar population in spiral arms of \\
      late type galaxies NGC~895, NGC~5474, and NGC~6946}

\author[A.~S.~Gusev, E.~V.~Shimanovskaya \& N.~A.~Zaitseva]
       {A.~S.~Gusev,\thanks{E-mail:gusev@sai.msu.ru}  
        E.~V.~Shimanovskaya 
        and N.~A.~Zaitseva \\
        Sternberg Astronomical Institute, Lomonosov Moscow State University, 
        Universitetsky pr. 13, 119234 Moscow, Russia \\
             }

\date{Accepted 2022 June 6. Received 2022 June 3; 
in original form 2022 April 20}

\begin{document}

\maketitle

\begin{abstract}
We investigate the spatial regularity in the distribution of the young 
stellar population along spiral arms of three late type spiral galaxies: 
NGC~895, NGC~5474, and NGC~6946. This study is based on an analysis of 
photometric properties of spiral arms using {\it GALEX} ultraviolet, optical 
$UBVRI$, H$\alpha$, and 8\,$\mu$m {\it IRAC} infrared surface photometry 
data. Using the Fourier analysis approach, we found features of spatial 
regularity or quasi-regularity in the distribution of the young stellar 
population or (and) regular chains of star formation regions in all arms 
of NGC~895, NGC~5474, and NGC~6946 with characteristic scales of spacing 
from 350 to 500~pc in different arms, and (or) scales which are multiples 
of them. These characteristic scales are close to the those found earlier 
in NGC~628, NGC 6217, and M100.
\end{abstract}


\section{Introduction}

Young stars are usually associated with spiral arms of disc 
galaxies. They are concentrated along the spiral arms non-uniformly 
and form hierarchical groupings of different scales 
\citep[see, e.g.,][]{efremov1998}. Sites of star formation are 
connected by unity of an origin with their ancestors -- the 
molecular clouds. Gravitational collapse and turbulence compression 
are suspected to play a key role in the creation and evolution of 
star formation regions. Spatial distribution of star formation 
regions in discs of spiral galaxies can be explained in terms of 
gravitational or magnetogravitational instability 
\citep[see, e.g.,][and references therein]{elmegreen1994,elmegreen2009}.

The regular spatial distribution of young stellar 
population in spiral arms is a rather rare phenomenon 
\citep{elmegreen1983}. Actually, theoretical studies of the 
gravitational instability of a gaseous 
\citep{safronov1960,elmegreen1983}, stellar-gaseous 
\citep{jog1984a,jog1984b,romeo2013}, and multicomponent 
disc \citep{rafikov2001} predict regularity in the 
distribution of molecular clouds and star formation 
regions at constant gas, $\Sigma_{\rm g}$, and stellar, 
$\Sigma_{\rm s}$, surface densities, gas sound speed, 
$c_{\rm g}$, and stellar velocity dispersion, $\sigma_{\rm s}$, 
on sufficiently large scales within a galactic disc.

The results of the study of the fragmentation of gas filaments 
\citep{inutsuka1997,matten2018} and multicomponent spiral arms 
\citep{inoue2018} also show the need for the constancy of the 
parameters of the stellar and interstellar medium along the 
filaments (spiral arms) for the formation of spatial regularities 
in distribution of star formation regions.

Note that theoretical calculations of the disc instability, as 
well as calculations of fragmentation of spiral arms, predict 
regularities on scales of a few kpc for typical values of 
parameters of the stellar and interstellar medium 
\citep{elmegreen1983,marchuk2018,inoue2021}.

The first studies on finding regularities seemed to confirm 
the theoretical models. \citet{elmegreen1983} found the regular 
strings of star complexes (H\,{\sc ii} regions) in spiral 
arms of 22 grand design galaxies with a characteristic spacing 
within 1-4~kpc. \citet{efremov2009,efremov2010} found the 
regularities in the distribution of H\,{\sc i} superclouds in 
spiral arms of our Galaxy and in the distribution of star 
complexes in M31. He estimated the average spacing between the 
H\,{\sc i} superclouds in the Carina spiral arm to be 1.5~kpc 
and in the Cygnus arm to be 1.3~kpc. He also found a regular 
string of star complexes in the north-western arm of M31 with 
a spacing of 1.1~kpc using {\it GALEX} UV images.

\begin{table*}
\caption[]{\label{table:sample}
The galaxy sample.
}
\begin{center}
\begin{tabular}{lcccccccccc} \hline \hline
Galaxy & Type & $B_t$ & $M_B^a$ & Inclination & PA       & 
$R_{25}^c$ & $R_{25}^c$ & $d$   & $A(B)_{\rm Gal}$ & $A(B)_{\rm in}$ \\
       &      & (mag) & (mag)   & (degree)    & (degree) & 
(arcmin)   & (kpc)      & (Mpc) & (mag)        & (mag) \\
1 & 2 & 3 & 4 & 5 & 6 & 7 & 8 & 9 & 10 & 11 \\
\hline
NGC~895  & SA(s)cd     & 12.30 & $-20.84$ & 
49 & 105 & 1.70 & 16.1 & 32.7 & 0.092 & 0.31 \\
NGC~5474 & SA(s)cd pec & 11.48 & $-18.07$ & 
50 & 100 & 1.20 &  2.3 &  6.7 & 0.038 & 0.33 \\
NGC~6946 & SAB(rs)cd   &  9.75 & $-20.68$ & 
31 &  62 & 7.74 & 13.3 &  5.9 & 1.241 & 0.04 \\
\hline
\end{tabular}
\end{center}
\begin{flushleft}
$^a$ Absolute magnitude of a galaxy corrected for Galactic extinction and 
inclination effects. \\
$^b$ Heliocentric radial velocity. \\
$^c$ Radius of a galaxy at the isophotal level 25 mag\,arcsec$^{-2}$ in 
the $B$ band corrected for Galactic extinction and inclination effects. \\
\end{flushleft}
\end{table*}

However, later there were indications of the presence of 
spatial regularities on sub-kpc scales. \citet{gusev2013} 
noted the existence of the same characteristic separation 
($\approx400$~pc) between adjacent star formation regions in 
both spiral arms of NGC~628. \citet*{elmegreen2018} found 
regularly spaced infrared peaks in the dusty spiral arms of 
M100 with a typical separation between them of 
$\approx410$~pc. Finally, \citet{gusev2020} estimated a 
regular characteristic spacing to be $\approx670$~pc between 
adjacent young stellar groupings in the ring of NGC~6217 with 
the suspicion an existence of characteristic separation with 
a half scale along the ring. \citet*{proshina2022} found 
a similar regular separation, $\sim700$~pc, for star formation 
regions in the ring of the lenticular galaxy NGC~4324.

\begin{figure}
\resizebox{1.00\hsize}{!}{\includegraphics[angle=000]{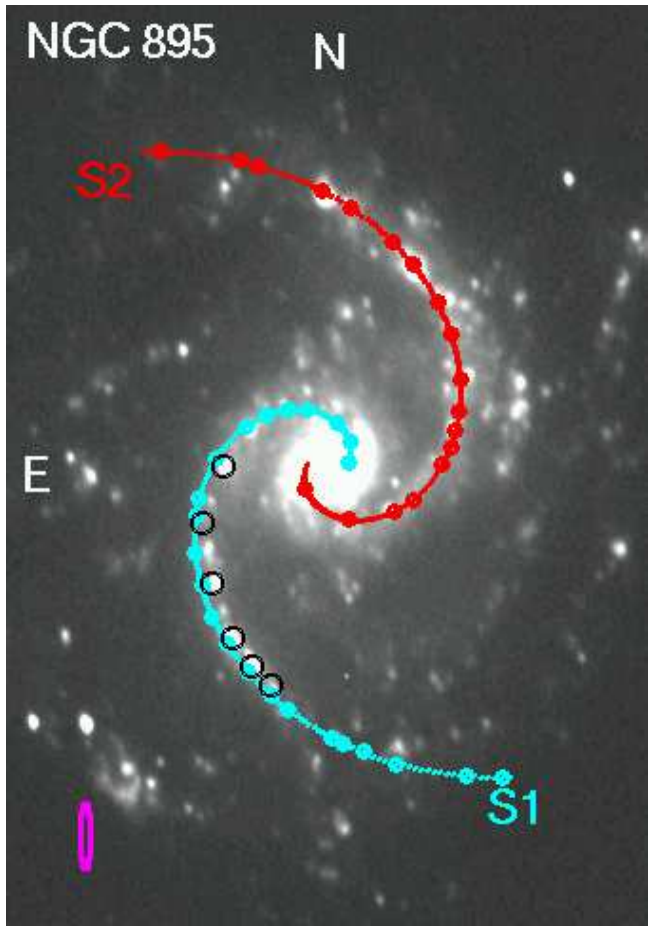}}
\caption{Deprojected image of NGC~895 in the H$\alpha$ line with 
overlaid logarithmic spirals S1 (cyan) and S2 (red). The cyan and red circles 
on the curves of the arms correspond to the positions of the local maxima of 
brightness. Positions of star formation complexes (H\,{\sc ii} regions) from 
the list of \citet{elmegreen1983} are indicated by open black circles. 
Example of used elliptical aperture ($18\times3$ arcsec$^2$) is shown 
in magenta. North is upward and east is to the left.
}
\label{figure:map895}
\end{figure}

The aim of this study is to further search for spatial regularities in the 
distribution of young stellar population in the spiral arms of galaxies and 
to estimate their characteristic spacing. We also want to understand how 
typical the spacing of $\sim0.5$~kpc found earlier in NGC~628 and M100?

For this study we selected three late type spiral galaxies: NGC~895, 
NGC~5474, and NGC~6946 (Table~\ref{table:sample}). Despite the similar 
morphological type, Scd, these galaxies have different structural features. 
NGC~895 is a classic grand gesign two-arms spiral 
(Fig.~\ref{figure:map895}). NGC~5474 is an asymmetric spiral galaxy with 
three arms located on one side of the centre (Fig.~\ref{figure:map5474}). 
Well known NGC~6946 is seem a rather flocculent (Fig.~\ref{figure:map6946}), 
however it has a regular spiral density wave \citep*{kendall2011,ghosh2016}. 
Two of the three galaxies, NGC~895 and NGC~5474, are on the list of 
\citet{elmegreen1983}.

NGC~5474 is about 10 times less massive and fainter than two 
other targets (see Table~\ref{table:sample}). It is similar to LMC in 
terms of mass and luminosity. The H\,{\sc i} mass in NGC~5474 is also 
approximately 10 times less than that in NGC~6946 \citep*{rownd1994,walter2008}, 
however, H$_2$-to-H\,{\sc i} mass ratio is equal to 0.86 in NGC~6946 
\citep{leroy2009}, but it does not reach 0.06 in NGC~5474 \citep{wilson2012}.

We use {\it GALEX} ultraviolet $FUV$ and $NUV$, optical $UBVRI$ and 
H$\alpha$ surface photometry data for the analysis of the spatial 
distribution of the young stellar population along the spiral arms 
of galaxies. Additionally, we use the available {\it IRAC} 8\,$\mu$m 
data for the NGC~6946.

The fundamental parameters of the galaxies are presented in
Table~\ref{table:sample}, where the morphological type, Galactic absorption, 
$A(B)_{\rm Gal}$, and the distance are taken from the 
NED\footnote{\url{http://ned.ipac.caltech.edu/}} database, and the remaining 
parameters are taken from the LEDA\footnote{\url{http://leda.univ-lyon1.fr/}} 
data base. The adopted value of the Hubble constant is equal to 
$H_0 = 75$~km\,s$^{-1}$Mpc$^{-1}$. With the assumed distances to the 
galaxies, we estimate their linear scales of 159, 32.5, and 
28.6~pc\,arcsec$^{-1}$ for NGC~895, NGC~5474, and NGC~6946, respectively.

\section{Observations and data reduction}

\subsection{Observational data}

Our own photometric $UBVRI$ and H$\alpha$+[N\,{\sc ii}] 
observations of NGC~6946 and data reduction were described 
earlier in \citet{gusev2016}.

Images of NGC~895 and NGC~5474 in H$\alpha$ line were 
downloaded via NED database. The image of NGC~895 was 
obtained in the {\it Survey for Ionization in Neutral Gas 
Galaxies (SINGG)}\footnote{\url{http://sungg.pha.jhu.edu/}} 
\citep{meurer2006}. The observations were carried out on 
December 30, 2000, with an exposure time of 1800~s. The 
resolution of the H$\alpha$ image is equal to 1.5~arcsec. 
H$\alpha$ image of NGC~5474 was obtained by \citet{knapen2004} on 
April 10, 2001, with an exposure of 1200~s and a seeing of 1.2~arcsec.

The absolute calibration of H$\alpha$ images were carried out 
using the descriptor parameters from the original FITS file 
and their explanations in \citet{meurer2006} and 
\citet{knapen2004}.

Ultraviolet {\it Galaxy Evolution Explorer (GALEX)} $FUV$ 
and $NUV$ reduced images of NGC~895, NGC~5474, and NGC~6946 were 
downloaded from the B.~A.~Miculski Archive for space
telescopes.\footnote{\url{http://galex.stsci.edu/}} The 
observations of NGC~895 were carried out on November 14, 2004, 
with a total exposure of 4869~s in every band. The 
observations of NGC~5474 were made on June 19, 2003, with a 
total exposure of 1610~s, the same in $FUV$ and $NUV$. The 
observations of NGC~6946 were carried out on September 9, 2006. 
The total exposure time was 894.35~s in $FUV$ filter and 
3208.65~s in $NUV$. The image resolution is equal to 4.5~arcsec 
for $FUV$ and 6.0~arcsec for $NUV$. The description of the 
{\it GALEX} mission, basic parameters 
of the passbands, files description, and data reduction were 
presented in \citet{morrissey2005}.

\begin{figure}
\resizebox{1.00\hsize}{!}{\includegraphics[angle=000]{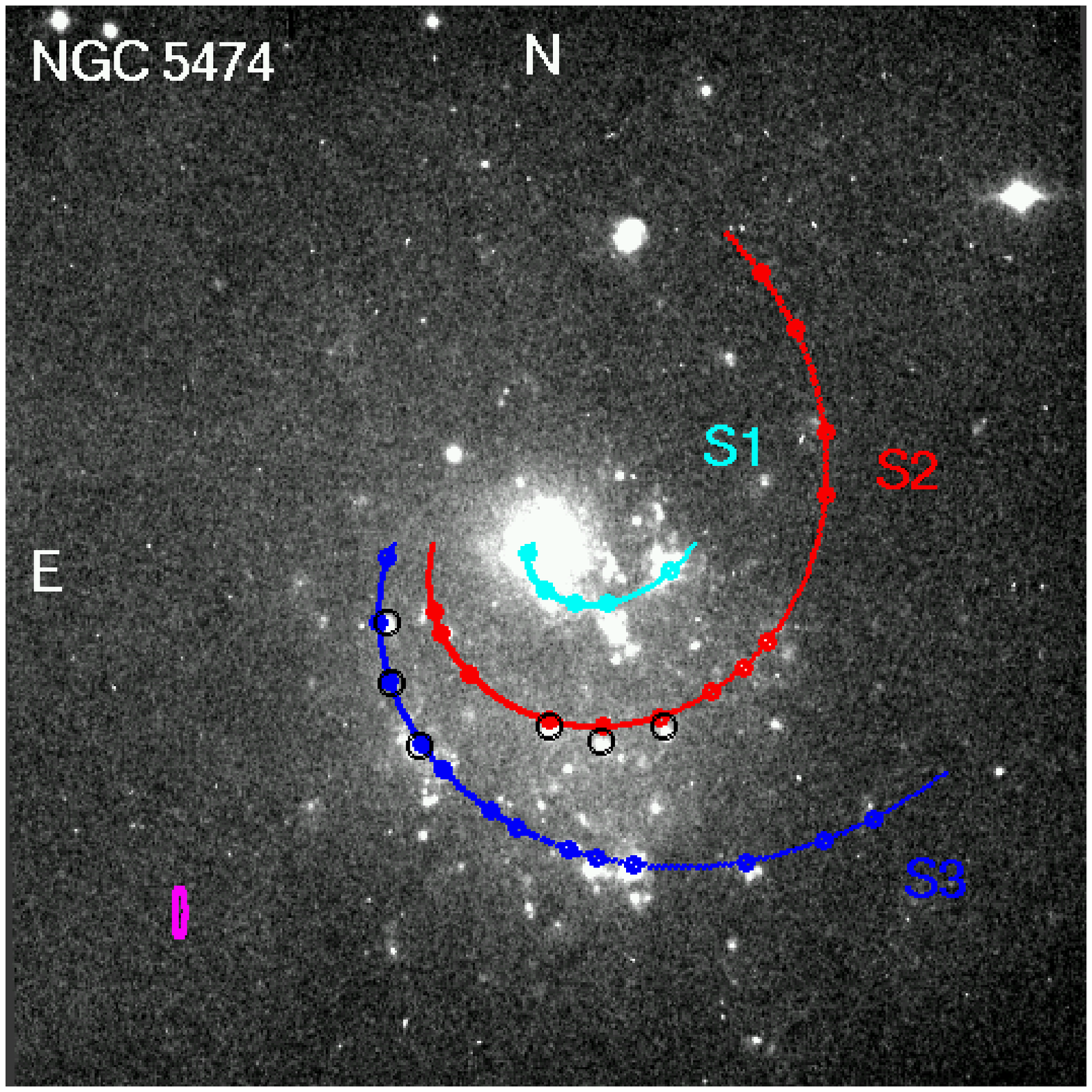}}
\caption{The same as Fig.~\ref{figure:map895}, but for NGC~5474. Logarithmic 
spiral S3 is shown in blue. The size of the used elliptical aperture is 
$18\times4$ arcsec$^2$.
}
\label{figure:map5474}
\end{figure}

\begin{figure}
\resizebox{1.00\hsize}{!}{\includegraphics[angle=000]{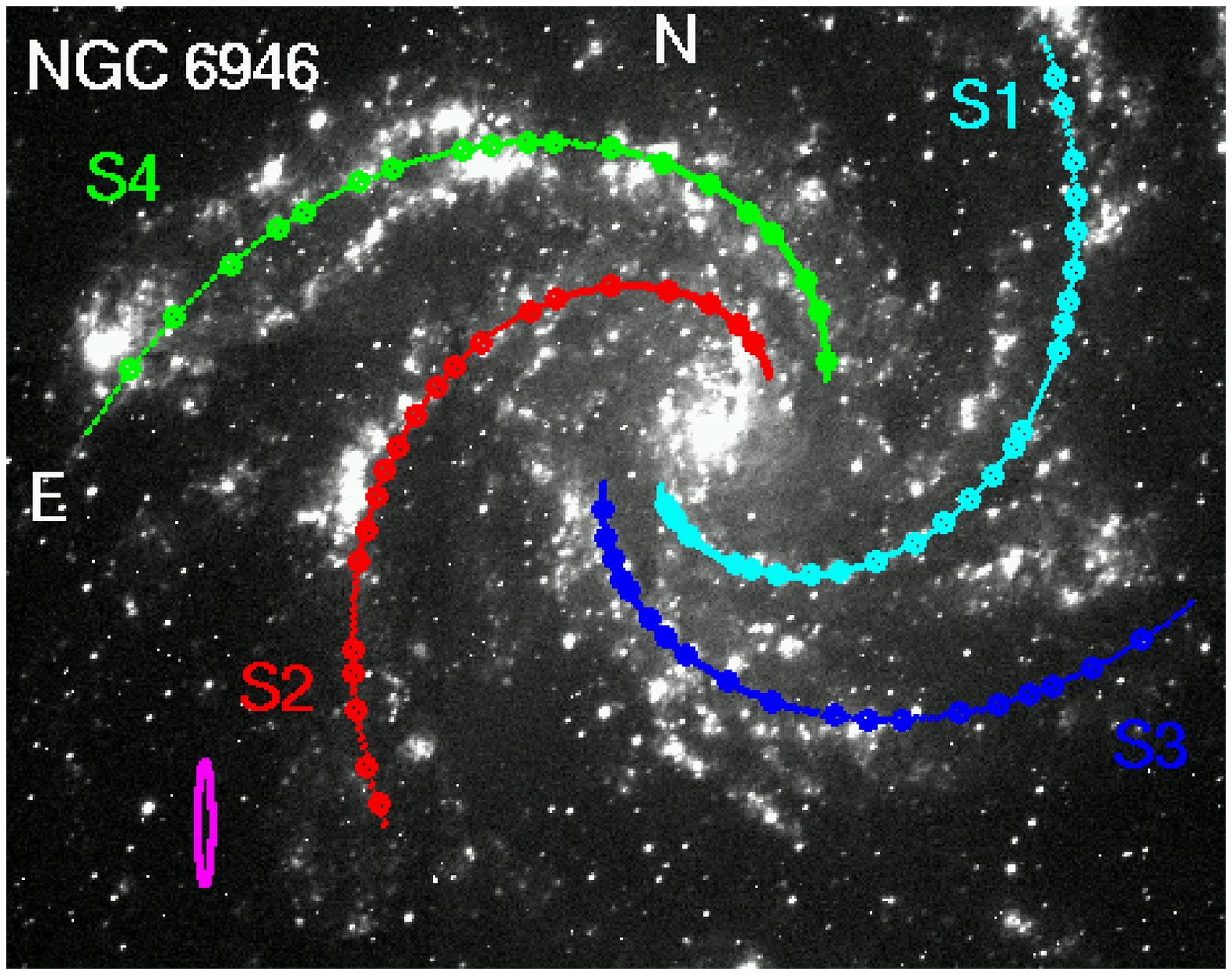}}
\caption{Same as Fig.~\ref{figure:map5474}, but for NGC~6946. Logarithmic 
spiral S4 is shown in green. The size of the used elliptical aperture is 
$50\times6$ arcsec$^2$.
}
\label{figure:map6946}
\end{figure}

The {\it Spitzer Space Telescope InfraRed Array Camera (IRAC)} image 
of NGC~6946 in the 8\,$\mu$m band was retrieved as a FITS file from the 
NED database. It was obtained in the {\it SIRTF Nearby Galaxies 
Survey (SINGS)}\footnote{\url{http://irsa.ipac.caltech.edu}} 
\citep{kennicutt2003}. The resolution of {\it IRAC} 8\,$\mu$m image 
is 2.4~arcsec. The observations of NGC~6946 were made on November 25, 
2004 with an exposure of 2144~s. Data reduction of {\it IRAC} images 
is described in the IRAC Instrument 
handbook\footnote{http://irsa.ipac.caltech.edu/data/SPITZER/docs/irac/ 

iracinstrumenthandbook/} 
\citep[see also][and references therein]{elmegreen2019}.

\subsection{Observations}

$UBVRI$ images of NGC~895 were obtained on October 10, 2004 
with the 1.5~m telescope of the Maidanak Observatory 
(Ulugh Beg Astronomical Institute of the Uzbekistan Academy of Sciences) 
using a SITe-2000 CCD array. The focal length of the telescope is 12~m. 
Detailed description of the telescope, the CCD camera, and the filters 
is presented in \citet{artamonov2010}. With broadband $U$, $B$, $V$, $R$, 
and $I$ filters, the CCD array realizes a photometric system close to the 
standard Johnson--Cousins $UBVRI$ system. The camera is cooled with 
liquid nitrogen. The size of the array is $2000\times800$ pixels. It 
provides the $8.9\times3.6$ arcmin$^2$ field of view with the image 
scale of 0.267 arcsec\,pixel$^{-1}$. The seeing during the observations 
was 0.9-1.3~arcsec. The total exposure times were 900~s in $U$ and $B$, 
720~s in $V$, and 540~s in $R$ and $I$.

Photometric $UBVRI$ observations of NGC~5474 were made on June 22, 2009 
with the same telescope with the SI-4000 CCD camera. The chip size, 
$4096\times4096$ pixels, provides a field of view of $18.1\times18.1$ 
arcmin$^2$ with an image scale of 0.267 arcsec\,pixel$^{-1}$. 
The total exposure times were 1200, 600, 420, 360, and 240~s in the 
$U$, $B$, $V$, $R$, and $I$ bands, respectively. The seeing was 
1.2-1.8~arcsec.

\subsection{Data reduction}

The reduction of the photometric data was carried out using 
standard techniques, with the European Southern Observatory
Munich Image Data Analysis System\footnote{\url
{http://www.eso.org/sci/software/esomidas/}} ({\sc eso-midas}). 
The main image reduction stages for photometric data were as follows: 
correction for bias and flat field, removal of cosmic-ray traces, 
determining and subtracting the sky background, aligning and adding the 
images, removal of field stars, absolute calibration. The main photometric 
$UBVRI$ image reduction stages were described in detail in 
\citet{gusev2016}.

The absolute calibration of the photometric data involved reducing 
the data from the instrumental photometric system to the standard 
Johnson--Cousins system and correcting for the air mass using the 
derived colour equations and the results of the aperture photometry 
of the galaxies.

For the absolute calibration of NGC~895 data we derived the colour 
equations and corrected for atmospheric extinction using observations 
of photometric standards from the field of \citet{landolt1992} SA~98, 
carried out on the same night and on the same air mass in the 
$U$, $B$, $V$, $R$, $I$ filters. A detailed study of the instrumental 
photometric system and the atmospheric extinction at the 
Maidanak Observatory was presented in \citet{artamonov2010}.

For the absolute calibration of NGC~5474 images the stars fields of 
\citet{landolt1992} PG~1657+078, SA~109, SA~110, and SA~111 were 
observed during the same set of observations in the wide air mass 
interval.

We also used the aperture photometric data for NGC~895 and NGC~5474 
from the LEDA database for the absolute calibration of images of the  
galaxies.

At the final step, the images of the galaxies in all bands were transformed 
to the face-on position using the values of the inclination ($i$) and the 
position angle (PA) from Table~\ref{table:sample}. Those face-on images 
were used for the further analysis. Photometric magnitudes and fluxes were 
corrected for the Galactic extinction and for projection effects:
$m_{\rm cor} = m_{\rm obs}-A_{\rm Gal}-2.5\log(\cos(i))$ and 
$I_{\rm cor} = I_{\rm obs}\,10^{0.4A_{\rm Gal}}\cos(i)$.

\section{Results}

\subsection{Parameters of spiral arms}

We investigate properties of spiral arms by curve fitting of spiral 
arms, which are clearly outlined in UV and blue optical ($U$, $B$) 
passbands (Figs.~\ref{figure:map895}-\ref{figure:map6946}). 
The arms are defined by by-eye selection of pixels in the part of 
face-on $B$ images that is within the spiral arms. Pixels in these 
regions are then fitted with a logarithmic spiral using linear 
least-squares.

A logarithmic spiral with a pitch angle $\mu$ is described as
\begin{equation}
r = r_0e^{k(\theta-\theta_0)},
\label{equation:spiral}
\end{equation}
where $k = \tan \mu$. We adopt two constraints for the fitting: 
(i) the pitch angle is constant along the arm, and (ii) the 
opposite spiral arms (S1 and S2 in NGC~895, S1 and S2, and S3 and S4 
in NGC~6946) have the same values of $\mu$ and $r_0$ 
(Figs.~\ref{figure:map895}, \ref{figure:map6946}). 

\begin{table}
\caption{Parameters of logarithmic spiral arms in the galaxies.}
\label{table:par_arms}
\centering
\begin{tabular}{rlccc}
\hline \hline
NGC & Spirals & $\mu$  & $r_0$    & $\theta_0$ \\
    &         & (degr) & (arcsec) & (degr) \\
\hline
895  & S1, S2 & $28.8\pm0.8$ & 5.75 & 270, 90 \\
5474 & S1     & $35.9\pm2.8$ & 5.62 & 90 \\
     & S2     & $15.3\pm1.0$ & 42.2 & 90 \\
     & S3     & $22.5\pm0.4$ & 57.2 & 90 \\
6946 & S1, S2 & $30.2\pm1.2$ & 32.0 & 135, 315 \\
     & S3, S4 & $31.0\pm1.0$ & 50.5 & 115, 295 \\
\hline
\end{tabular}
\end{table}

Results of the fitting are presented in Table~\ref{table:par_arms}. 
Remark that the obtained pitch angles for the arms in NGC~6946, 
$\mu = 30.2\degr$ and $31.0\degr$, are close to the results of 
\citet*{kendall2015} who obtained values $24.0\degr\pm0.6\degr$, 
$29.3\degr\pm0.5\degr$, and $29.5\degr\pm0.7\degr$ using optical $V$, 
3.6\,$\mu$m, and 4.5\,$\mu$m images, respectively.

\subsection{Along-arm photometry}

For the further analysis, we used a technique developed by us in 
\citet{gusev2013} and \citet{gusev2020}. To study brightness
variations along spiral arms, we obtained photometric profiles of them.
We used elliptical apertures with a minor axis along a spiral arm 
(a difference between PA of major axis and PA of the centre of 
aperture is a pitch angle), and a step of $1\degr$ by PA. The step 
used corresponds to the angular distance from 0.1 to 2~arcsec for 
spiral arms in NGC~895, from 0.1 to 3~arcsec for spiral arms in 
NGC~5474, and 0.6 to 4~arcsec for spiral arms in NGC~6946; it does 
not exceed the angular resolution in the $FUV$ band.

The sizes of the apertures, $18\times3$ arcsec$^2$ for NGC~895, 
$18\times4$ arcsec$^2$ for NGC~5474, and $50\times6$ arcsec$^2$ for 
NGC~6946, were selected taking into account the width of the spiral 
arms and characteristic sizes of star formation regions. The major 
axis was sized to cover all star formation regions in the spiral arms, 
but exclude regions outside of them. The size of the minor axis was 
chosen so that adjacent star formation regions were separated in 
photometric profiles. These apertures are plotted in 
Figs.~\ref{figure:map895}-\ref{figure:map6946}.

\begin{figure}
\vspace{0.4cm}
\resizebox{1.00\hsize}{!}{\includegraphics[angle=000]{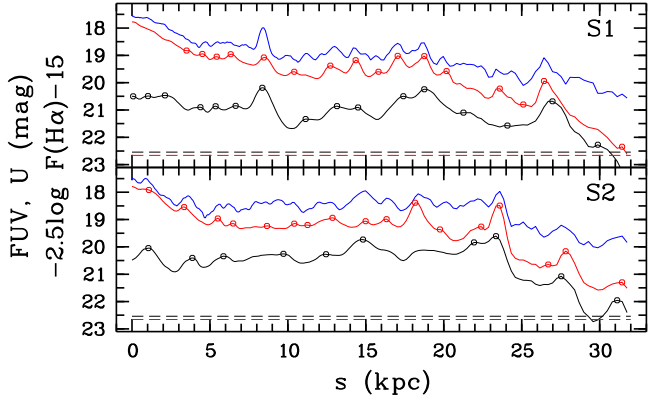}}
\caption{Photometric profiles in the $U$ (blue), $FUV$ (black) 
bands, and H$\alpha$ line (red) along the spiral arms S1 (top) and S2 (bottom) 
of NGC~895. The ordinate units are magnitudes and logarithm of H$\alpha$ flux 
in units of erg\,s$^{-1}$cm$^{-2}$ within the aperture. Positions of local 
maxima of brightness on the profiles in the $FUV$ band (open black circles) 
and H$\alpha$ line (open red circles) are indicated. 3$\sigma$ thresholds 
above the average background level for $FUV$ (black) and 
H$\alpha$ images (red) are shown by dashed horizontal lines.
}
\label{figure:profile895}
\end{figure}

\begin{figure}
\vspace{0.3cm}
\resizebox{1.00\hsize}{!}{\includegraphics[angle=000]{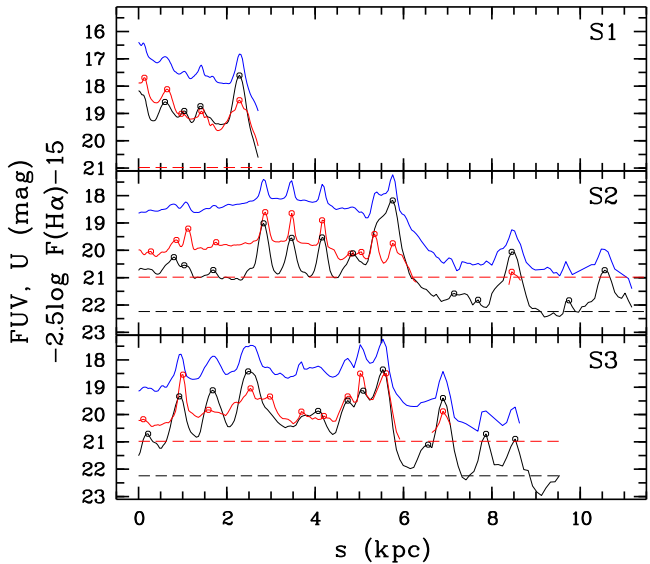}}
\caption{Photometric profiles in the $U$ (blue), $FUV$ (black) 
bands, and H$\alpha$ line (red) along the spiral arms S1 (top), S2 (middle), 
and S3 (bottom) of NGC~5474. Other symbols are the same as in 
Fig.~\ref{figure:profile895}.
}
\label{figure:profile5474}
\end{figure}

\begin{figure}
\vspace{0.4cm}
\resizebox{1.00\hsize}{!}{\includegraphics[angle=000]{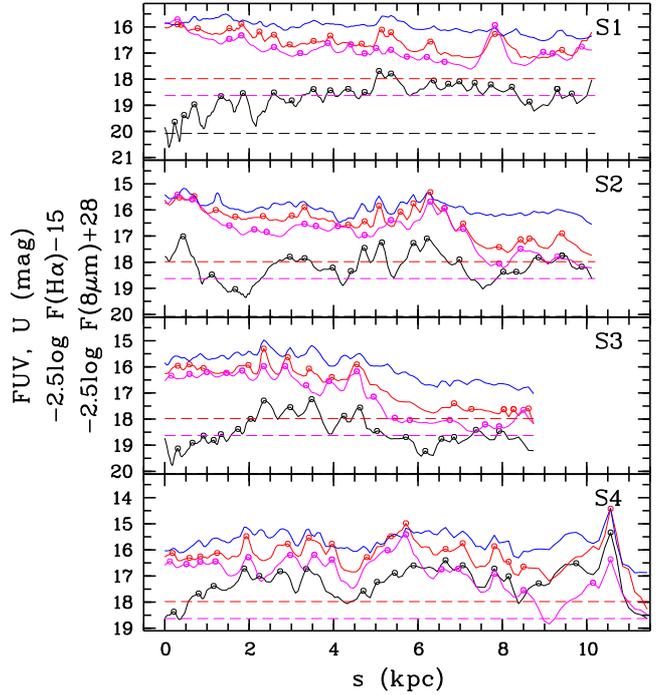}}
\caption{Photometric profiles in the $U$ (blue), $FUV$ (black) 
bands, H$\alpha$ line (red), and 8\,$\mu$m (magenta) along the spiral arms 
S1--S4 of NGC~6946. 8\,$\mu$m flux is given in units of Jy within the 
aperture. Positions of local maxima of brightness on the profiles in the 
8\,$\mu$m IRAC band are indicated by open magenta circles. The
3$\sigma$ threshold above the average background level for the 8\,$\mu$m image 
is shown by the magenta dashed horizontal line. Other symbols are the same as 
in Fig.~\ref{figure:profile895}.
}
\label{figure:profile6946}
\end{figure}

Obtained photometric profiles in $FUV$, $U$, and H$\alpha$ along the 
arms of the galaxies are presented in 
Figs.~\ref{figure:profile895}-\ref{figure:profile6946}, where the 
longitudinal displacement along the spiral, denoted as $s$, is
\begin{equation}
s = (\sin \mu)^{-1}r_0(e^{k(\theta-\theta_0)}-1)
\label{equation:sp_long}
\end{equation}
for a logarithmic spiral in the form of Eq.~(\ref{equation:spiral}). 
Additionally, we present 8\,$\mu$m profiles along the spiral arms of 
NGC~6946 in Fig.~\ref{figure:profile6946}.

Using the profiles in the $FUV$ band and H$\alpha$ line, and involving 
profiles in the $U$ band, we found the local brightness maxima on the 
profiles. We prefer to use the $FUV$ and H$\alpha$ images that show 
the distribution of the young stellar population with ages of 1-10~Myr 
(H$\alpha$) and 10-100~Myr ($FUV$). The $U$ images were used for control.

The local maxima of brightness were determined as the lower extrema of the 
functions $m_{FUV}(s)$, $-\log F({\rm H}\alpha)(s)$ for the spiral arms. 
To locate them, we looked for points with the first derivative of the 
function, $dm_{FUV}/ds = 0$ or $d(\log F({\rm H}\alpha))/ds = 0$, and 
the second derivative, $d^2m_{FUV}/ds^2 > 0$ or 
$d^2(-\log F({\rm H}\alpha))/ds^2 > 0$, on the profiles. We selected 
peaks whose widths exceed three measurement points 
(corresponding to the angular resolution of the images in optical bands) 
and whose amplitude exceeds the $3\sigma$ threshold above the average 
background level in the corresponding images of the galaxies within the 
used apertures (see dashed horizontal lines and open circles in 
Figs.~\ref{figure:profile895}-\ref{figure:profile6946}).

The outer parts of the spiral arms S2 and S3 in NGC~5474 are faintly 
visible in the H$\alpha$ line. Due to the low signal-to-noise ratio, we 
excluded sections of profiles with $\log F({\rm H}\alpha)<-14.4$ from 
further consideration. The only $FUV$ and $U$ profiles were used for 
these sections (Fig.~\ref{figure:profile5474}).

The local maxima of brightness found in both $FUV$ and H$\alpha$ were 
taken as a single local maximum of brightness if the difference between their 
positions along the spiral arm, $|s_{FUV}-s_{{\rm H}\alpha}|$, was less 
than the FWHM of the $FUV$ peak profile and less than the FWHM of the 
H$\alpha$ peak profile.

\begin{figure}
\vspace{4mm}
\resizebox{1.00\hsize}{!}{\includegraphics[angle=000]{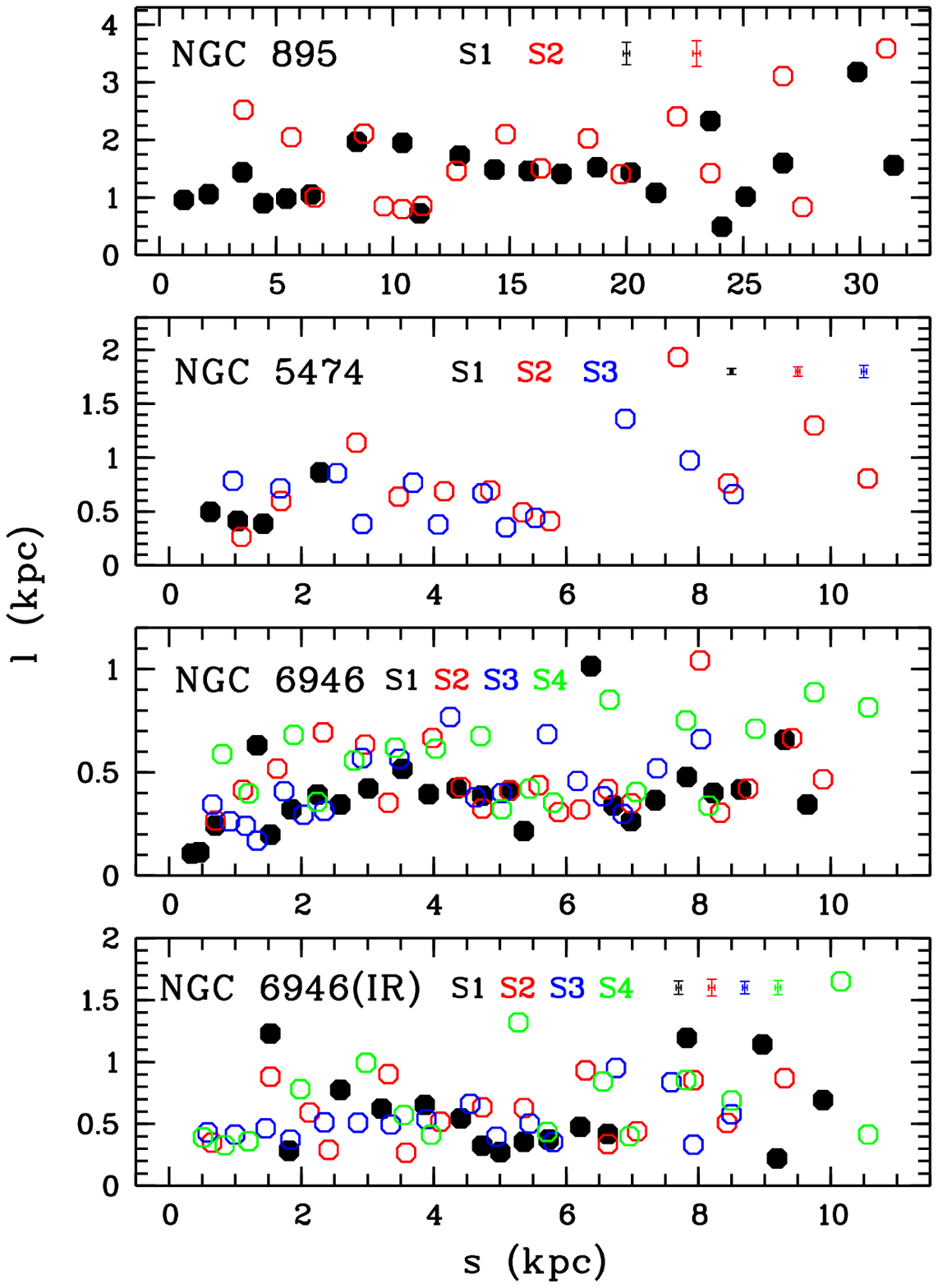}}
\caption{Separations $l$ between adjacent local maxima of brightness 
along the spiral arms of NGC~895, NGC~5474, and NGC~6946 from the 
UV and optical data (three top panels) and along the spiral arms of 
NGC~6946 from the IR data (bottom panel). Separations $l$ along 
arms S1--S4 are indicated by black, open red, open blue, and open 
green circles, respectively. The mean error bars are shown. See the 
text for details.
}
\label{figure:separate}
\end{figure}

The search for local maxima of brightness using the infrared 8\,$\mu$m 
profiles of the spiral arms in NGC~6946 (Fig.~\ref{figure:profile6946}) 
was carried out separately using the same algorithm.

As can be seen from Figs.~\ref{figure:profile895}-\ref{figure:profile6946}, 
in a number of cases, the local maxima of brightness found from the 
$FUV$ and H$\alpha$ profiles do not coincide with each other. This is a 
consequence of the fact that young stellar populations of different 
ages radiate in different wavelength ranges.

As a result, we obtained 23 local maxima of brightness in the spiral arm 
S1 and 18 local maxima of brightness in the spiral arm S2 of NGC~895 
(Fig.~\ref{figure:profile895}); 5, 13, and 13 local maxima of brightness 
in the spiral arms S1, S2, and S3 of NGC~5474, respectively 
(Fig.~\ref{figure:profile5474}); 25, 21, 19, and 19 local maxima of 
brightness in the spiral arms S1, S2, S3, and S4 of NGC~6946, 
respectively (Fig.~\ref{figure:profile6946}). Analysis of infrared 
8\,$\mu$m profiles of NGC~6946 gave 17 local maxima of brightness for 
the spiral arms S1 and S3, and 16 local maxima of brightness for the 
spiral arms S2 and S4 (Fig.~\ref{figure:profile6946}).

\begin{figure}
\vspace{3mm}
\resizebox{1.00\hsize}{!}{\includegraphics[angle=000]{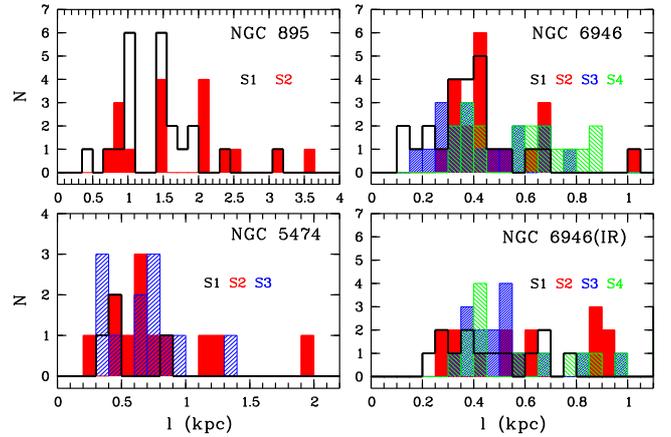}}
\caption{The number distribution histograms of local maxima of brightness 
by separation between adjacent objects along the spiral arms of NGC~895 
(top-left), NGC~5474 (bottom-left), NGC~6946 (top-right) obtained from the 
UV and optical data, and along the spiral arms of NGC~6946 obtained from the 
IR data (bottom-right). The colours of the curves corresponding to the 
various spiral arms are the same as in Fig.~\ref{figure:separate}. See the 
text for more explanation.
}
\label{figure:hist}
\end{figure}

The typical errors of maximum positions are $\Delta s = 1$ 
measurement points. Taking into account the linear resolution of the 
H$\alpha$ images of galaxies, mean positions errors are equal to 
$\pm150$~pc for NGC~895, $\pm20$, $\pm30$, and $\pm40$~pc for the arms 
S1, S2, and S3 of NGC~5474, respectively, $\pm40$~pc for the spirals S1, 
S2, and S3 in NGC~6946, and $\pm50$~pc for the spiral arm S4 of 
NGC~6946 (see error bars in Fig.~\ref{figure:separate}). Individual 
uncertainties for the identified local maxima of brightness 
are presented in Table~\ref{table:data}.

In the next step, we measured separations, $l$, between adjacent 
local maxima of brightness along the spiral arms in the galaxies. 
Separation, $l$, between ($n-1$)-st and $n$-th local maxima of 
brightness is defined as $l_n = s_n-s_{n-1}$, where $s \equiv s_n$ is 
defined from Eq.~(\ref{equation:sp_long}). Distribution of the 
local maxima of brightness by separation $l$ in the arms of galaxies is 
presented in Fig.~\ref{figure:separate}. Numerical values of $s$, $l$, and 
galactocentric distances $r$ for the local maxima of brightness and 
their pairs are presented in Table~\ref{table:data}.

The obtained separations are larger, with a few exceptions, than 
a linear resolution of the $FUV$ images (700~pc for NGC~895, 150~pc for 
NGC~5474, and 130~pc for NGC~6946). Several measurements with smaller 
separations are confirmed by H$\alpha$ profiles along the spiral arms. 
Linear resolutions of the H$\alpha$ images (240~pc for NGC~895, 40~pc 
for NGC~5474, and 30~pc for~NGC 6946), as well as 8\,$\mu$m image 
linear resolution for NGC~6946 (70~pc) are significantly smaller than 
minimum values of $l$ (see Figs.~\ref{figure:separate}, 
~\ref{figure:hist}).

As we noted above, FWHMs of the peak profiles exceed the linear 
resolutions of the H$\alpha$ and optical images of the galaxies in all 
cases.

Photometric profiles along spiral arms in 
Figs.~\ref{figure:profile895}-\ref{figure:profile6946} cannot be used
directly for estimating photometric parameters 
of individual young star clusters and complexes because of highly 
elliptical apertures used to obtain them. Several star clusters and 
H\,{\sc ii}~regions may be found to be in the measured area due to the large 
major axis. On the other hand, the size of the minor axis, 3-6~arcsec, 
is smaller than angular sizes of large star complexes. Peak H$\alpha$ 
fluxes and $FUV$ magnitudes may be used for measuring 
photometric fluxes of star formation regions only for compact 
separate H\,{\sc ii}~regions and star clusters.

We estimated star formation rates of star formation 
regions in the galaxies using their H$\alpha$ luminosities and $FUV$ 
absolute magnitudes with the conversion factor of H$\alpha$ luminosity to 
star formation rate of \citet{kennicutt1998}:
\begin{eqnarray}
{\rm SFR (}M_\odot \,{\rm yr^{-1})} = 
7.9\cdot10^{-42} L({\rm H}\alpha) {\rm (erg\,s^{-1})} \nonumber
\end{eqnarray}
and the conversion factor of $FUV$ luminosity to star formation rate of 
\citet{iglesias2006} in the form 
\begin{eqnarray}
{\rm SFR (}M_\odot \,{\rm yr^{-1})} = 
7.0\cdot10^{-8}\cdot10^{-0.4M_{FUV}}. \nonumber
\end{eqnarray}
Photometric luminosities and fluxes were corrected for the extinction 
due inclination effects (see Table~\ref{table:sample}).

We roughly (to the order of magnitude) estimated masses of the young 
star clusters, assuming that a character time scale of star formation 
in OB associations and star complexes is $\sim10$~Myr 
\citep{efremov1998}. More precise mass estimates require more 
detailed analysis of photometric and spectral parameters and are 
beyond the scope of this study.

\subsection{Analysis of the spatial distribution of 
local maxima of brightness}

We built histograms of the distribution of separations $l$ in the 
spiral arms of the galaxies (Fig.~\ref{figure:hist}). In addition, 
we decided to perform the Fourier analysis to search for 
characteristic separations between local maxima of brightness.

Observational data are often noisy and unevenly spaced both in the 
time domain and in the spatial domain. Therefore the direct use of 
the Fourier analysis to investigate periodicities in those data can 
be rather problematic. The normalized periodogram was initially defined 
by \citet{scargle1982} to estimate the power spectrum for unevenly 
sampled time series. We use the modified periodogram method of 
\citet{horne1986} for spatial data series of separations $l$ between 
local maxima of brightness in the spiral arms of the galaxies. 
To estimate the significance of the height of a peak in the power 
spectrum, the concept of the false alarm probability (FAP) was 
adopted \citep{scargle1982}. The FAP characterizes the probability 
that a peak of a certain height or higher will occur, assuming that 
the data are pure noise. An important parameter for the calculation 
of the FAP is the number of independent frequencies $N$ from 
$\omega_0 = 2\pi/L$ to $\omega_N = \pi N_0/L$, where $L$ is the 
largest separation, $N_0$ is the number of data points in the initial 
data series. Through simulation of a large number of data sets, 
\citet{horne1986} empirically derived the formula for $N$: 
\begin{eqnarray} 
N = -6.362 + 1.193N_0 + 0.00098N_0^2. \nonumber
\end{eqnarray}
This formula is accurate for evenly spaced data but it overestimates 
the number of independent frequencies for unevenly sampled data. 
Therefore false alarm probabilities for the unevenly sampled data, 
as in our study, could be significantly smaller.

So we computed the Lomb-Scargle periodograms, using the technique 
developed by \citet{scargle1982}, \citet{horne1986}, and \citet{press1989}, 
and also thoroughly explained in \citet{vp2018}, for our data to estimate 
the power spectrum. For this, we used functions $D(s)$,  $D_n(s)$, 
$p(s)$, and $p_n(s)$. The function $D(s) = 1$ in points of the local 
maximum of brightness and $D(s) = 0$ in all other points. The function 
$D_n(s)$ is a function $D(s)$ normalized to the maximum amplitude of 
$FUV$ or H$\alpha$ peak. The function $p(s)$ is a collection of Gaussians, 
centered at points of local maxima of brightness on the profiles, with 
$\sigma$ equal to the peak positioning error. The function $p_n(s)$ is a 
function $p(s)$ with the Gaussian amplitudes normalized to the maximum 
amplitude. If significant peaks were recorded in both bands ($FUV$ and 
H$\alpha$), the largest normalized amplitude was adopted. Outside of the 
Gaussians, the functions $p(s)$ and $p_n(s) = 0$ for all other points.

\begin{figure}
\vspace{4mm}
\resizebox{1.00\hsize}{!}{\includegraphics[angle=000]{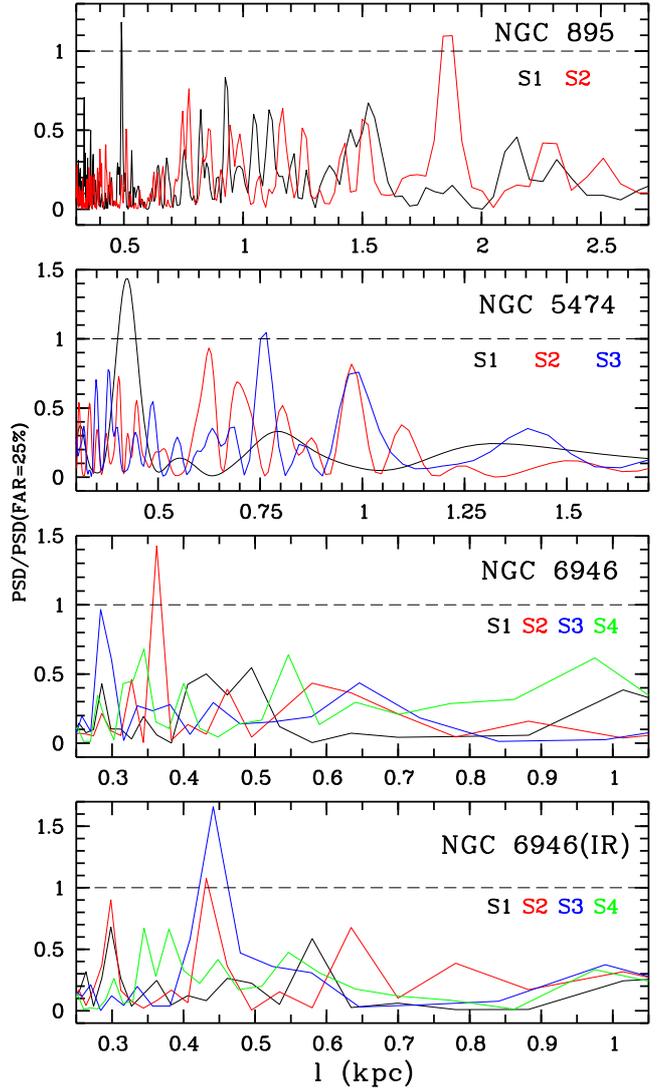}}
\caption{Normalized power spectral density of the function $p(s)$ for the 
spiral arms in NGC~895, NGC~5474, and NGC~6946 from the UV and optical data, 
and for the spiral arms of NGC~6946 from the IR data. The colours of the 
curves corresponding to the various spiral arms are the same as in 
Fig.~\ref{figure:separate}. See the text for more explanation.
}
\label{figure:power}
\end{figure}

An analysis of the periodograms for the functions $D(s)$, $D_n(s)$, 
$p(s)$, $p_n(s)$ showed that they all have peaks at the same $l$ (for each 
individual spiral arm), but with slightly different power spectral density 
(PSD) values. In Fig.~\ref{figure:power} we present the periodograms for 
the function $p(s)$. 

In \citet{gusev2013} and \citet{gusev2020} 
we showed that the Fourier analysis of raw intensity profiles reveals only  
long-scale regularities. It is not sensitive to the short-scale regularities 
which are of interest to us. So we do not show periodograms calculated 
directly for unmodified deprojected intensities along the arms
in $FUV$ and H$\alpha$ in Fig.~\ref{figure:power}.

Because we have different numbers of local maxima of brightness 
$N_0$ for different spiral arms in NGC~895, NGC~5474, and NGC~6946, 
numbers of independent frequencies $N$ differ too, and therefore 
normalized periodograms in Fig.~\ref{figure:power} have different smoothness. 

Power spectral density (ordinate axis) in Fig.~\ref{figure:power} 
is normalized to the PSD with a false-alarm probability 
\citep[FAP; see][for details]{horne1986} equal to $25\%$.

Below, we consider the distribution of star formation regions along 
the spiral arms separately for each spiral arm.

\subsubsection{NGC 895}

For NGC 895, \citet{elmegreen1983} found a regular chain of 
H\,{\sc ii} regions in S1, consisting of six objects with a separation of 
about 1.4~kpc (Fig.~\ref{figure:map895}). All regions of \citet{elmegreen1983} 
are identified by us as local maxima of brightness. They are located at the 
longitudinal displacement $s=8-20$~kpc (galactocentric distances 
$r=4.8-10.5$~kpc; see Figs.~\ref{figure:profile895}, 
\ref{figure:separate}). Figure~\ref{figure:separate} shows a rather regular 
separation between adjacent local maxima of brightness along the S1 with 
separations $l\approx1$~kpc at $s=0-7$~kpc ($r=0.9-4.3$~kpc) 
and $l\approx1.5$~kpc at $s=13-20$~kpc ($r=7.2-10.5$~kpc). The 
difference between separations found by \citet{elmegreen1983} 
and ourselves is explained, firstly, by different adopted values of 
the distance and inclination of the disc of NGC~895, and, secondly, by 
the fact that we found more local maxima of brightness than 
\citet{elmegreen1983} in the same segment of the spiral arm 
(see Fig.~\ref{figure:map895}). We also note that all distances 
between adjacent local maxima of brightness are a multiple of 0.5~kpc 
with an accuracy of $\pm10\%$. The exception is a trio of regions at 
$s=11-13$~kpc ($r=6.2-7.2$~kpc) with distances of 0.7 and 1.7~kpc and 
a pair of regions at $s=23.5$~kpc ($r=12.2$~kpc) with a distance of 
2.3~kpc (Fig.~\ref{figure:separate}).

Distribution of star formation regions in the S1 by separations have two 
clear peaks at $l\approx1$~kpc and $l\ge1.5$~kpc 
(Fig.~\ref{figure:hist}). The distribution of separations can be bimodal. 
Given the sample size is too small to make a robust multimodality 
test, we calculated distances $D$ between the peaks relative to their 
widths according to the definition by \citet*{ashman1994}:
$D=\vert \mu_a-\mu_b \vert / \sqrt{(\sigma_a^2+\sigma_b^2)/2}$, where 
$\mu_a$, $\mu_b$ are the means of subsets $a$ and $b$, $\sigma_a$, 
$\sigma_b$ are the standard deviations, which include the intrinsic 
measurement errors in pairs, $\pm200$~pc. As is known, $D>2$ is required for 
a clean separation between the modes. We selected two subsets. The first one
includes seven pairs with separations from 0.9 to 1.1~kpc, and the second one
includes eight pairs with separations from 1.4 to 1.6~kpc. The mean 
separations in the subsets are $1.01\pm0.20$~kpc and $1.49\pm0.22$~kpc, 
which gives $D=2.32$ and testifies in favour of a bimodal distribution of 
separations.

The periodogram shows a noticeable peak with a FAP $<20\%$ at 
$l=490$~pc for the S1 (Fig.~\ref{figure:power}). Weak additional peaks at 
separations $\approx1$~kpc and $\approx1.5$~kpc have a FAP $>30\%$. 
But we should remember that the significance of a peak in the periodogram 
is underestimated in the case of unevenly sampled data. 

\citet{elmegreen1983} did not find any regularity in distribution 
of H\,{\sc ii} regions in the spiral arm S2 of NGC~895. Our analysis shows 
the presence of regularities with scales of $\sim1$, $\sim$~1.5, and 
$\sim2$~kpc, including a chain of four star formation regions with 
$l=840$~pc at $s=9-12$~kpc ($r=5.2-6.7$~kpc; Fig.~\ref{figure:separate}).

These three characteristic scales are confirmed by the corresponding 
peaks in the histogram in Fig.~\ref{figure:hist}. The mean separations 
in these three subsets, corresponding to the pairs with separations of 
$\sim1$, $\sim1.5$, and $\sim2$~kpc, are $0.87\pm0.23$, $1.45\pm0.21$, 
and $2.07\pm0.20$~kpc, including the intrinsic errors. The parameter $D$ is 
equal to 2.65 for subsets with $l\sim1$ and $\sim1.5$~kpc, and 3.02 for 
subsets with $l\sim1.5$ and $\sim2$~kpc. At the same time, the periodogram 
has a single significant peak with a FAP $<25\%$ at $l=1.85-1.9$~kpc 
(Fig.~\ref{figure:power}). We remark that the characteristic separations, 
found in the S2, are multiples of 500~pc, as in the S1 arm.

The brightest and largest H\,{\sc ii}~regions in the galaxy are located 
in S2 at $s=18$ and 24~kpc (Fig.~\ref{figure:profile895}). Their sizes, 
600-650~pc, are typical for star complexes. Luminosities of these complexes 
in H$\alpha$ reach $9.1\cdot10^{39}$~erg\,s$^{-1}$ which corresponds to 
SFR $=0.07 M_\odot\,{\rm yr^{-1}}$ and $m\sim7\cdot10^5 M_\odot$. The brightest 
H\,{\sc ii}~regions in the opposite arm, S1, at $s=8.5$ and 18~kpc 
(Fig.~\ref{figure:profile895}) are fainter. Their H$\alpha$ luminosity 
is equal to $4.0\cdot10^{39}$~erg\,s$^{-1}$, 
SFR $=0.03 M_\odot\,{\rm yr^{-1}}$, and $m\sim3\cdot10^5 M_\odot$. 
The faintest H\,{\sc ii}~regions, detected in the profiles in 
Fig.~\ref{figure:profile895} in the galaxy, have 
$L({\rm H}\alpha)=1.2\cdot10^{38}$~erg\,s$^{-1}$, 
SFR $=0.001 M_\odot\,{\rm yr^{-1}}$, and $m\sim9\cdot10^3 M_\odot$.

The brightest young star complexes in $FUV$ are also 
systematically brighter in S2 arm than in S1. The two brightest complexes 
with $M_{FUV}=-14.40$~mag are located at $s=15$ and 24~kpc 
(Fig.~\ref{figure:profile895}). Their SFRs and masses, 
$0.07 M_\odot\,{\rm yr^{-1}}$ and $\sim7\cdot10^5 M_{\odot}$, turned out 
to be the same as those calculated from the H$\alpha$ luminosity. We note 
that the star complex at $s=15$~kpc, which is the brightest in the $FUV$, 
is not distinguished by the H$\alpha$ emission 
(Fig.~\ref{figure:profile895}).

In contrast to the sample of star formation regions in S2, all star 
complexes in S1, that are bright in $FUV$, are powerful H\,{\sc ii}~regions. 
The estimates of SFRs, $0.016 M_\odot\,{\rm yr^{-1}}$, and masses, 
$\sim1.6\cdot10^5 M_{\odot}$, of the brightest star complexes in $FUV$ in 
S1 with $M_{FUV}=-12.76$~mag are slightly smaller than ones obtained 
from the H$\alpha$ data.

The faintest young star clusters, visible in $FUV$ band, have
$M_{FUV}\approx-11.1$~mag, which corresponds to 
SFR $=0.0035 M_\odot\,{\rm yr^{-1}}$ and $m\sim3.5\cdot10^4 M_{\odot}$. 
Higher minimal SFR detected in $FUV$ than in H$\alpha$ is due to the 
lower resolution and signal-to-noise ratio of $FUV$ observations.

Masses of the brightest star complexes in NGC~895 are close to 
masses of the largest complexes in giant late type galaxies, such 
as M51, where the mass of the largest star formation complex is 
$\sim3\cdot10^5 M_{\odot}$ \citep{larsen2001}, and NGC~6946, with 
the most massive stellar complex of $\sim8\cdot10^5 M_{\odot}$ 
\citep{bastian2005}.

\subsubsection{NGC 5474}

Only five local maxima of brightness were found in the 
inner short spiral arm of NGC~5474 (S1; see 
Figs.~\ref{figure:map5474}, \ref{figure:profile5474}). The first 
four of them form a regular chain with a separation $430\pm60$~pc, 
and the distance between the fourth and fifth local maxima of 
brightness is twice as large and equal to $860\pm30$~pc 
(Figs.~\ref{figure:separate}, \ref{figure:hist}). The Fourier analysis 
of a sample of five objects can only be formal; it shows the 
selected scale of 420~pc with a FAP $<15\%$ (Fig.~\ref{figure:power}).

Distribution of star formation regions in S2 looks rather irregular 
(Figs.~\ref{figure:separate}, \ref{figure:hist}). Some regularity 
seems to be observed in the inner part of this spiral arm at $s=1.5-5$~kpc 
($r=1.8-2.7$~kpc), where we found six local maxima of brightness. Five of 
them are located at a distance $660\pm60$~pc from each other, and one 
pair has a double separation $l=1.14\pm0.05$~kpc 
(Fig.~\ref{figure:separate}). The four most distant star formation 
regions from this sample were identified by \citet{elmegreen1983} as 
a regular chain of four H\,{\sc ii} regions 
(see Fig.~\ref{figure:map5474}).

Power spectral density of the function $p(s)$ for S2 never reaches 
the level of $25\%$ FAP. The PSD maximum at $l\approx620$~pc probably 
reflects the fact of quasi-regularity in distribution of star formation 
regions at $s=1.5-5$~kpc in the S2 arm (Fig.~\ref{figure:power}).

Unlike the S2, the spiral arm S3 in the galaxy shows a regular 
distribution of the young stellar population along the arm.
Two characteristic separations are stand out in the S3. We found four pairs 
of adjacent local maxima of brightness with a separation of 
$l=390\pm70$~pc and six pairs with a spacing of $l=740\pm90$~pc 
(Figs.~\ref{figure:separate}, \ref{figure:hist}). The bimodal separation 
distribution is confirmed by the value of the parameter $D=4.34$. 
The spectrum of the function $p(s)$ for S3 also confirms the 
characterictic distance of 750~pc with a FAP $=25\%$ 
(Fig.~\ref{figure:power}). This scale was noted in \citet{elmegreen1983}.

Note that the H$\alpha$-to-$FUV$ surface brightness ratio  
along the spiral arms is systematically lower in NGC~5474 than in NGC~895 
and NGC~6946 (compare 
Figs.~\ref{figure:profile895}-\ref{figure:profile6946}). This is especially 
noticeable for arms S1 and S3 (Fig.~\ref{figure:profile5474}). Apparently, 
this is a consequence of the relative deficit of H$_2$ in the 
less-massive NGC~5474.

The sizes of the largest H\,{\sc ii}~regions, observed in the galaxy, 
are 200-250~pc. The brightest star formation regions are located in 
the inner spiral arm S1 at $s=1.5$ and 2.4~kpc 
(Fig.~\ref{figure:profile5474}). Their 
$L({\rm H}\alpha)=5.1\cdot10^{38}$~erg\,s$^{-1}$ and $M_{FUV}=-12.72$~mag. 
These values correspond to SFR $=0.004 M_\odot\,{\rm yr^{-1}}$ and 
$m\sim4\cdot10^4 M_{\odot}$ by H$\alpha$ luminosity and 
SFR $=0.009 M_\odot\,{\rm yr^{-1}}$ and $m\sim9\cdot10^4 M_{\odot}$ by 
$FUV$ absolute magnitude.

Outer spiral arms S2 and S3 contain star formation regions with 
$L({\rm H}\alpha)\le3.0\cdot10^{38}$~erg\,s$^{-1}$ and 
$M_{FUV}\ge-11.80$~mag which correspond to 
SFR $=0.002-0.004 M_\odot\,{\rm yr^{-1}}$ and 
$m\sim(2-4)\cdot10^4 M_{\odot}$. The brightest ones are located at 
$s=2.9$, 3.5, 4.1, and 5.5~kpc in S2, and at $s=5.0$ and 5.5 kpc in S3 
(Fig.~\ref{figure:profile5474}). Note that the bright $FUV$ source at 
$s=5.5$~kpc in S2 does not emit in H$\alpha$. In spite of the similar 
SFRs (masses) of the brightest star formation regions in S2 and S3 arms, 
young regions in S2 are more compact. Their diameters are $\le150$~pc.

Smaller maximal masses of star formation regions in NGC~5474 as 
compared with NGC~895 and NGC~6946 are a consequence of the smaller 
total mass of NGC~5474 and the smaller fraction of the molecular hydrogen in 
it as compared to the giant galaxies NGC~895 and NGC~6946.

The faintest star formation regions, derived from the $FUV$ and H$\alpha$ 
images, have $L({\rm H}\alpha)\approx1.6\cdot10^{37}$~erg\,s$^{-1}$ 
and $M_{FUV}\approx-9.0$~mag. SFR in them is equal to 
$1\cdot10^{-4} M_\odot\,{\rm yr^{-1}}$ ($m\sim1\cdot10^3 M_{\odot}$) 
based on H$\alpha$ data and $2.5\cdot10^{-4} M_\odot\,{\rm yr^{-1}}$ 
($m\sim3\cdot10^3 M_{\odot}$) based on $FUV$ luminosity.

\subsubsection{NGC 6946}

Most of spiral arms in NGC~6946 look ragged. The only exception 
is the spiral arm S2 (Fig.~\ref{figure:map6946}). It would be hard 
to expect an accurate regularity in distribution of star formation regions 
here.

Spatial distribution of local maxima of brightness in the spiral arm S1 
does not show a certain characteristic separation. However, most 
of the pairs, 13 out of 24, have separations in a narrow range of 
320--420~pc (Figs.~\ref{figure:separate}, \ref{figure:hist}). Among 
them, we can distinguish a regular chain of five star formation 
regions at $s=3.5-5.2$~kpc ($r=2.7-3.5$~kpc) with a spacing of 
$400\pm60$~pc (Fig.~\ref{figure:separate}).

The distribution of infra-red sources in S1 looks even more chaotic, 
although all distances between adjacent regions lie in the same 
range of separations as found in optics and UV (see right panels in 
Fig.~\ref{figure:hist}).

Local maxima of brightness, detected in UV and H$\alpha$, are distributed 
in the spiral arm S2 rather regularly. 13 pairs out of 20 have separations 
from 310 to 470~pc with a mean $l=380\pm85$~pc 
(Figs.~\ref{figure:separate}, \ref{figure:hist}). This scale is confirmed 
by the Fourier analysis: the periodogram shows a strong peak 
at $l=360$~pc with a FAP $<15\%$ (Fig.~\ref{figure:power}).

Distribution of regions of incipient star formation visible in 8\,$\mu$m 
shows stronger spatial regularity in S2 than the distribution of star 
formation regions visible in UV and H$\alpha$ line 
(Fig.~\ref{figure:separate}). Three  separations were found 
for pairs of IR sources: $310\pm65$, $555\pm95$, and 
$890\pm70$~pc (Fig.~\ref{figure:hist}). Segregation of the subsets is 
confirmed by the value of parameter $D$, which is equal to 3.01 for subsets 
with $l\sim300$ and $l\sim550$~pc, and is equal to 4.01 for subsets 
with $l\sim550$ and $l\sim900$~pc. Note that both larger scales are 
multiples of the smallest scale 300~pc.

Power spectral density of the function $p(s)$ for IR sources in S2 
reaches its maximum at $l=300$ and 430~pc with a FAP $\approx25\%$ 
(Fig.~\ref{figure:power}). The first maximum coincides with the smallest 
scale. The second maximum lies between scales of $\sim300$ and 
$\sim550$~pc. The discrepancy between the scales obtained using the 
Fourier analysis and those found from the histogram is apparently due 
to the absence of regular chains of star formation regions in S2. 
There are no adjacent pairs with the same separation in this arm 
(Fig.~\ref{figure:separate}).

Distribution of star formation regions, visible in optics and UV in the 
spiral arm S3, is similar to those in the inner arms S1 and S2 
(Fig.~\ref{figure:hist}). More than a half of pairs have separations from 
240 to 410~pc with a mean $l=330\pm80$~pc. This spacing scale is 
similar to the result of the Fourier analysis: PSD reaches a maximum 
at 280~pc with a FAP $\approx25\%$ (Fig.~\ref{figure:power}).

We observe two regular chains of IR sources in the inner part of S3. 
The first four pairs at $s=0.5-2$~kpc ($r=1.2-2.4$~kpc) have a 
separation $l=420\pm60$~pc, and the second four pairs at 
$s=2.4-4$~kpc ($r=2.7-3.5$~kpc) have a separation $l=515\pm50$~pc 
(Fig.~\ref{figure:separate}). The regularity in the distribution of 
IR sources is lost in the outer parts of the spiral arm. In general, 
two characteristic scales of separations can be distinguished along 
the spiral arm, seven pairs with $l$ in the range of $330-460$~pc and 
six pairs with spacings from 500 to 580~pc (Fig.~\ref{figure:hist}). 
However, these subgroups do not segregated by the criterion of 
\citet{ashman1994}: with average values of $395\pm70$ and $520\pm65$~pc, a 
parameter $D=1.86$. Note that, as in the case of the spiral arm S2 of the 
galaxy, the periodogram shows a significant peak with a FAP $<15\%$ at 
$l=440$~pc, in the middle of two subsets (Fig.~\ref{figure:power}).

Histograms in Fig.~\ref{figure:hist} have peaks in the distribution of star 
formation regions by separation in the brightest spiral arm of NGC~6946, 
S4, at $l\approx400$~pc for both UV-H$\alpha$ and IR sources. The 
distances between seven pairs of adjacent UV and H$\alpha$ emission 
regions lie in a narrow range of $320-420$~pc with a mean
$l=370\pm75$~pc. All other pairs have separations $l\ge560$~pc. Among 
the wide pairs, we suspect a slow increase in the typical separation $l$ 
with the galactocentric distance $r$ (see Fig.~\ref{figure:separate}).

The distribution of IR sources along the S4 arm completely coincides with 
the distribution of UV-H$\alpha$ regions. Seven close pairs with 
separations of $330-430$~pc have a mean $l=390\pm70$~pc. Wide pairs 
have spacings $l\ge570$~pc. At the same time, the periodograms 
do not show any significant peaks at $300-400$~pc (see 
Fig.~\ref{figure:power}).

As in NGC~895, we observe an asymmetry of distribution 
of young stellar population in the disc of NGC~6946. The star formation is 
more active in the eastern spiral arms S2 and S4 
(Fig.~\ref{figure:map6946}). The brightest source of modern star formation 
is located on the end of S4 at $s=10.5$~kpc (Fig.~\ref{figure:profile6946}). 
It has $L({\rm H}\alpha)=6.8\cdot10^{39}$~erg\,s$^{-1}$ and 
$M_{FUV}=-13.79$~mag, which corresponds to 
SFR $=0.054 M_\odot\,{\rm yr^{-1}}$ from H$\alpha$ data and 0.023 from $FUV$ 
data. The mass of this region, $\sim(2-5)\cdot10^5 M_{\odot}$, is close 
to the mass of the brightest star complex in NGC~895.

The brightest star formation regions in S2, located in a chain at 
$s=4.5-7.0$~kpc (Fig.~\ref{figure:profile6946}), are approximately twice as 
faint and less massive than those in S4. The brightest star formation regions 
in the western spiral arms S1 at $s=5.1$~kpc and S3 at $s=2.4-4.6$~kpc 
reach H$\alpha$ luminosities $1.4\cdot10^{39}$~erg\,s$^{-1}$ and $FUV$ 
absolute magnitudes $-11.20$. Their SFR, estimated using H$\alpha$ and 
$FUV$ data, are equal to $0.010 M_\odot\,{\rm yr^{-1}}$ 
($m\sim1\cdot10^5 M_{\odot}$) and $0.002 M_\odot\,{\rm yr^{-1}}$ 
($m\sim2\cdot10^4 M_{\odot}$), respectively.

The faintest star clusters and H\,{\sc ii}~regions, detectable in the galaxy 
with $L({\rm H}\alpha)\approx4.2\cdot10^{37}$~erg\,s$^{-1}$ and 
$M_{FUV}\approx-9.9$~mag, have SFR $=4\cdot10^{-4} M_\odot\,{\rm yr^{-1}}$ 
and $m\sim4\cdot10^3 M_{\odot}$ from H$\alpha$ measurements and 
SFR $=7\cdot10^{-4} M_\odot\,{\rm yr^{-1}}$ and $m\sim7\cdot10^3 M_{\odot}$ 
from $FUV$ measurements.

The most massive star complex in the galaxy, with a diameter of 600~pc and 
a mass of up to $8\cdot10^5 M_{\odot}$ \citep{bastian2005}, is located in 
the disc in the middle of S1 and the end of S3; therefore, it was not 
included in our study. This is the brightest $FUV$ source 
($M_{FUV}=-14.06$~mag), but in the H$\alpha$ line it looks like a group of 
weak H\,{\sc ii}~regions located along the edges of the complex 
(Fig.~\ref{figure:map6946}). This is due to its age, 15~Myr 
\citep{bastian2005}.

\section{Discussion}

Our analysis of the spatial distribution showed that in most 
spiral arms of NGC~895 and NGC~5474 there is a regularity in the 
distribution of star formation regions along the spiral arms of the 
galaxies. In the arm S2 of NGC 5474, where no characteristic regularity 
scale has been found, there is a regular chain consisting of several star 
formation regions. In the case of NGC~6946, we could rather talk about a 
quasi-regularity in the spatial distribution of star formation regions. 
The spacings between adjacent regions lie in a narrow range; however, 
regular chains are not observed in all spiral arms, except for S2.

Thus, the presence of a characteristic scale of distances between 
adjacent star formation regions along spiral arms seems to be more 
common than previously thought. The absence of strong regularity in 
three of four spiral arms of NGC~6946 can be explained by their 
morphology. In S1 and S4, the parallel chains of star formation regions 
are observed along the outer and inner edges of the spiral arms; spiral 
arm S3 branches into several segments (Fig.~\ref{figure:map6946}). Such 
structures might blur a possible spatial scale in the distribution 
of star formation regions.

We note that in spiral arms where several characteristic 
separations are found, the larger ones are multiples of the smallest 
one. At the same time, the smallest separations $l\sim350-500$~pc in 
most spiral arms are close to the characteristic spacings 
obtained earlier for NGC~628, NGC~6217, and M100 
\citep{gusev2013,elmegreen2018,gusev2020}. In the distant galaxy NGC~895, 
the characteristic separations are multiples of $\sim500$~pc, however, 
our angular resolution does not allow us to study the galaxy at this 
scale. We encountered a similar situation earlier when analysing the spatial 
regularity in the distant galaxy NGC~6217 \citep{gusev2020}.

We attempted to apply the criterion of the gravitational instability of the
gaseous, stellar-gaseous and multicomponent disc 
\citep{safronov1960,elmegreen1983,jog1984a,jog1984b,romeo2013,rafikov2001} 
for numerical calculations of the observed scale of regularities in the 
distribution of the young stellar population in spiral arms. Calculations 
of the perturbation wavelength in the disc of NGC~628, in which regular 
chains of star formation regions with a characteristic spacing of 
$\approx400$~pc are observed \citep{gusev2013}, and in the disc of 
NGC~6946 based on modern data on surface density and velocity dispersions 
of atomic and molecular hydrogen and of the disc stellar population 
\citep[projects 
{\it THINGS}\footnote{\url{http://www.mpia.de/THINGS}} \citep{walter2008}, 
{\it BIMA SONG}\footnote{\url{http://adil.ncsa.uiuc.edu/document/02.TH.01}} 
\citep{helfer2003}, {\it HERACLES} \citep{leroy2009}, {\it SINGS} 
\citep{kennicutt2003}; see the detailed technique in][]{leroy2008} 
predict characteristic separations between adjacent star formation 
regions of the order of several kiloparsecs, provided that the parameters 
are averaged over scales larger than the instability wavelength. This is 
well illustrated by the instability wavelength map in NGC~628 
\citep[fig.~4 in][]{marchuk2018}. At the same time, an approximate 
constancy of the surface density and velocity dispersion of atomic and 
molecular hydrogen and of the disc stellar population is indeed observed 
in NGC~628 and 
NGC~6946 in the range of galactocentric distances, where regularity 
in the distribution of star formation regions was found. Approximate 
constancy of the H\,{\sc i} density is also observed in a wide range 
of galactocentric distances in NGC~5474 \citep{rownd1994,pascale2021}. 
As we noted in Introduction, fragmentation calculations for gaseous 
filaments and multicomponent spiral arms 
\citep{inutsuka1997,matten2018,inoue2018} also predict a regularity 
scale larger than 1~kpc for NGC~628 \citep{inoue2021}. Thus, current 
theoretical approaches cannot adequately explain the sub-kiloparsec 
regularity scale 
in the distribution of the young stellar population along the spiral arms.

We have not found any noticeable difference in the 
distribution of star formation regions with an age of $3-100$~Myr, 
visible in the UV and optics, and younger star formation regions, 
visible in the IR, in NGC~6946.

A magnetic field in a galaxy can change instability conditions and 
parameters, such as instability growth time and wavelength 
\citep{elmegreen1983}. It can stabilize and freeze the initial spatial 
regularities, if they exist. For example, a regular magnetic field with 
a wavelength of 2.3~kpc was found by \citet{beck1989} in the same arm 
segment of M31 in which \citet{efremov2009,efremov2010} noticed a regular 
string of star complexes with a spacing of 1.1~kpc.

\citet{frick2000} found that the magnetic arms in NGC~6946, visible in 
polarized radio emission at 3.5 and 6.2~cm, are localized almost 
precisely between the optical arms; and each magnetic arm is similar 
in length and pitch angle to the preceding optical arm. The regions of 
maximum polarized intensity in the galaxy are located near the inner parts 
of optical spiral arms S1 and S2, and near the middle part of the S3 arm. 
It is in these parts of the spiral arms that we have not found regular 
chains of star formation regions (see Fig.~\ref{figure:separate}). 
On the other hand, the ratio of the excess in the regular magnetic field 
over the azimuthal average
at 3.5~cm in the arms turns out to be maximum 
at galactocentric distances of $3-6$~kpc 
\citep[see fig.~7 in][]{frick2000}. It is at these galactocentric 
distances, that regular chains of star formation regions, consisting of 
$3-5$ objects, are observed in most spiral arms of NGC~6946.

\section{Conclusions}

We found a spatial regularity or quasi-regularity in the distribution of 
star formation regions along arms and (or) regular chains of star 
formation regions consisting of $3-5$ objects in all spiral arms of 
galaxies studied in this paper. It is worth noting that these three 
galaxies have noticeably different structural features.

The obtained characteristic separations between adjacent star 
formation regions are from 350 to 500~pc or (and) multiples of them in 
the majority of spiral arms. The larger characteristic separations are 
multiples of the smallest one in those spiral arms where several 
characteristic separations are observed.

The characteristic scales of $350-500$~pc obtained in NGC~895, NGC~5474, 
and NGC~6946 are close to those found earlier in NGC~628, M100, 
and NGC~6217.

\section*{Acknowledgments}
We are grateful to the anonymous referee for his/her constructive 
comments. The authors acknowledge the use of the HyperLeda data base 
(\url{http://leda.univ-lyon1.fr}), the NASA/IPAC Extragalactic 
Database (\url{http://ned.ipac.caltech.edu}), Barbara~A.~Miculski Archive 
for space telescopes (\url{https://archive.stsci.edu}), 
the H\,{\sc i} Nearby Galaxy Survey (THINGS) data archive 
(\url{http://www.mpia.de/THINGS}) and the IDL Astronomy 
User's Library (\url{https://idlastro.gsfc.nasa.gov}). The authors 
would like to thank {\it SINGG}, {\it GALEX}, {\it SINGS}, {\it THINGS}, 
{\it BIMA SONG}, {\it HERACLES}, and the project of \citet{knapen2004} 
teams for making observational data available. This study was supported by 
the Russian Foundation for Basic Research (project no. 20-02-00080).

\section*{Data availability}
The {\it GALEX} UV data used in this paper are available in the 
Barbara~A.~Miculski Archive for space telescopes at 
\url{https://galex.stsci.edu}. The H$\alpha$ data for NGC~895 
and NGC~5474, and 8\,$\mu$m {\it IRAC} data for NGC~6946 are available 
in NASA/IPAC Extragalactic Database at \url{http://ned.ipac.caltech.edu}. 
The $UBVRI$ data for the galaxies and H$\alpha$ data for NGC~6946 
can be shared on reasonable request to the corresponding author.

\appendix

\section{Positions of local maxima of brightness along spiral 
         arms and separations between adjacent local maxima of brightness}

\begin{table}
\caption{Galactocentric distances, longitudinal displacements along 
spiral arms, and separations for local maxima of brightness and 
their pairs.}
\label{table:data}
\centering
\begin{tabular}{rrrr}
\hline \hline
$n$ & $r_n$ (kpc) & \multicolumn{1}{c}{$s_n$ (kpc)} & 
\multicolumn{1}{c}{$l_n$ (kpc)} \\
\hline
\multicolumn{4}{c}{NGC 895 -- S1} \\
 2 &  1.42 &  1.05$\pm$0.12 & 0.96$\pm$0.17 \\
 3 &  1.93 &  2.11$\pm$0.12 & 1.06$\pm$0.17 \\
 4 &  2.62 &  3.55$\pm$0.12 & 1.44$\pm$0.17 \\
 5 &  3.06 &  4.45$\pm$0.12 & 0.90$\pm$0.17 \\
 6 &  3.53 &  5.43$\pm$0.12 & 0.98$\pm$0.17 \\
 7 &  4.04 &  6.49$\pm$0.12 & 1.05$\pm$0.17 \\
 8 &  4.98 &  8.46$\pm$0.12 & 1.97$\pm$0.17 \\
 9 &  5.92 & 10.41$\pm$0.12 & 1.95$\pm$0.17 \\
10 &  6.27 & 11.13$\pm$0.12 & 0.73$\pm$0.17 \\
11 &  7.11 & 12.86$\pm$0.14 & 1.73$\pm$0.20 \\
12 &  7.82 & 14.35$\pm$0.12 & 1.48$\pm$0.17 \\
13 &  8.53 & 15.81$\pm$0.12 & 1.46$\pm$0.17 \\
14 &  9.20 & 17.22$\pm$0.18 & 1.41$\pm$0.26 \\
15 &  9.94 & 18.75$\pm$0.12 & 1.52$\pm$0.17 \\
16 & 10.63 & 20.18$\pm$0.12 & 1.43$\pm$0.17 \\
17 & 11.15 & 21.26$\pm$0.22 & 1.08$\pm$0.31 \\
18 & 12.27 & 23.59$\pm$0.12 & 2.33$\pm$0.17 \\
19 & 12.51 & 24.09$\pm$0.12 & 0.49$\pm$0.18 \\
20 & 13.00 & 25.10$\pm$0.13 & 1.02$\pm$0.18 \\
21 & 13.77 & 26.70$\pm$0.27 & 1.60$\pm$0.39 \\
22 & 15.30 & 29.88$\pm$0.15 & 3.18$\pm$0.21 \\
23 & 16.05 & 31.44$\pm$0.16 & 1.56$\pm$0.23 \\
\hline
\multicolumn{4}{c}{NGC 895 -- S2} \\
 2 &  2.65 &  3.60$\pm$0.12 & 2.52$\pm$0.17 \\
 3 &  3.63 &  5.65$\pm$0.12 & 2.05$\pm$0.17 \\
 4 &  4.11 &  6.65$\pm$0.12 & 1.00$\pm$0.17 \\
 5 &  5.13 &  8.76$\pm$0.12 & 2.11$\pm$0.17 \\
 6 &  5.54 &  9.61$\pm$0.12 & 0.85$\pm$0.17 \\
 7 &  5.92 & 10.40$\pm$0.12 & 0.80$\pm$0.17 \\
 8 &  6.33 & 11.26$\pm$0.12 & 0.85$\pm$0.17 \\
 9 &  7.04 & 12.72$\pm$0.14 & 1.46$\pm$0.20 \\
10 &  8.05 & 14.82$\pm$0.12 & 2.10$\pm$0.17 \\
11 &  8.77 & 16.33$\pm$0.12 & 1.51$\pm$0.17 \\
12 &  9.75 & 18.36$\pm$0.19 & 2.03$\pm$0.27 \\
13 & 10.43 & 19.76$\pm$0.21 & 1.41$\pm$0.29 \\
14 & 11.59 & 22.17$\pm$0.23 & 2.41$\pm$0.33 \\
15 & 12.27 & 23.59$\pm$0.12 & 1.42$\pm$0.17 \\
16 & 13.77 & 26.70$\pm$0.27 & 3.11$\pm$0.39 \\
17 & 14.17 & 27.53$\pm$0.28 & 0.83$\pm$0.40 \\
18 & 15.89 & 31.12$\pm$0.16 & 3.59$\pm$0.22 \\
\hline
\multicolumn{4}{c}{NGC 5474 -- S1} \\
2 & 0.55 & 0.62$\pm$0.02 & 0.50$\pm$0.03 \\
3 & 0.79 & 1.04$\pm$0.02 & 0.41$\pm$0.03 \\
4 & 1.02 & 1.42$\pm$0.02 & 0.39$\pm$0.03 \\
5 & 1.52 & 2.29$\pm$0.02 & 0.86$\pm$0.03 \\
\hline
\multicolumn{4}{c}{NGC 5474 -- S2} \\
 2 & 1.66 &  1.09$\pm$0.03 & 0.27$\pm$0.04 \\
 3 & 1.82 &  1.69$\pm$0.02 & 0.60$\pm$0.03 \\
 4 & 2.12 &  2.83$\pm$0.04 & 1.14$\pm$0.05 \\
 5 & 2.29 &  3.47$\pm$0.02 & 0.64$\pm$0.03 \\
 6 & 2.47 &  4.16$\pm$0.02 & 0.69$\pm$0.03 \\
 7 & 2.66 &  4.85$\pm$0.02 & 0.69$\pm$0.03 \\
 8 & 2.78 &  5.35$\pm$0.02 & 0.49$\pm$0.04 \\
 9 & 2.89 &  5.76$\pm$0.03 & 0.41$\pm$0.04 \\
10 & 3.41 &  7.69$\pm$0.06 & 1.93$\pm$0.09 \\
11 & 3.61 &  8.45$\pm$0.03 & 0.76$\pm$0.05 \\
12 & 3.95 &  9.75$\pm$0.04 & 1.30$\pm$0.05 \\
13 & 4.16 & 10.56$\pm$0.04 & 0.81$\pm$0.05 \\
\hline
\end{tabular}
\end{table}

\setcounter{table}{0}
\begin{table}
\caption{Continued}
\centering
\begin{tabular}{rrrr}
\hline \hline
$n$ & $r_n$ (kpc) & \multicolumn{1}{c}{$s_n$ (kpc)} &
\multicolumn{1}{c}{$l_n$ (kpc)} \\
\hline
\multicolumn{4}{c}{NGC 5474 -- S3} \\
 2 & 2.23 & 0.96$\pm$0.04 & 0.79$\pm$0.06 \\
 3 & 2.51 & 1.68$\pm$0.05 & 0.72$\pm$0.07 \\
 4 & 2.83 & 2.54$\pm$0.03 & 0.86$\pm$0.04 \\
 5 & 2.98 & 2.92$\pm$0.06 & 0.38$\pm$0.08 \\
 6 & 3.27 & 3.69$\pm$0.03 & 0.77$\pm$0.04 \\
 7 & 3.42 & 4.07$\pm$0.03 & 0.38$\pm$0.04 \\
 8 & 3.68 & 4.74$\pm$0.03 & 0.67$\pm$0.05 \\
 9 & 3.81 & 5.09$\pm$0.04 & 0.35$\pm$0.05 \\
10 & 3.98 & 5.53$\pm$0.04 & 0.44$\pm$0.05 \\
11 & 4.50 & 6.89$\pm$0.04 & 1.36$\pm$0.06 \\
12 & 4.88 & 7.87$\pm$0.05 & 0.97$\pm$0.06 \\
13 & 5.13 & 8.53$\pm$0.05 & 0.66$\pm$0.07 \\
\hline
\multicolumn{4}{c}{NGC 6946 -- S1} \\
 2 & 1.09 & 0.34$\pm$0.02 & 0.11$\pm$0.02 \\
 3 & 1.15 & 0.46$\pm$0.02 & 0.11$\pm$0.02 \\
 4 & 1.27 & 0.70$\pm$0.03 & 0.24$\pm$0.04 \\
 5 & 1.59 & 1.33$\pm$0.02 & 0.63$\pm$0.02 \\
 6 & 1.69 & 1.53$\pm$0.02 & 0.20$\pm$0.02 \\
 7 & 1.85 & 1.85$\pm$0.02 & 0.32$\pm$0.03 \\
 8 & 2.04 & 2.24$\pm$0.02 & 0.39$\pm$0.03 \\
 9 & 2.22 & 2.59$\pm$0.02 & 0.34$\pm$0.03 \\
10 & 2.43 & 3.01$\pm$0.02 & 0.42$\pm$0.03 \\
11 & 2.69 & 3.53$\pm$0.03 & 0.52$\pm$0.04 \\
12 & 2.89 & 3.92$\pm$0.03 & 0.39$\pm$0.04 \\
13 & 3.10 & 4.35$\pm$0.03 & 0.42$\pm$0.04 \\
14 & 3.30 & 4.73$\pm$0.03 & 0.39$\pm$0.05 \\
15 & 3.51 & 5.15$\pm$0.07 & 0.41$\pm$0.10 \\
16 & 3.62 & 5.36$\pm$0.04 & 0.22$\pm$0.05 \\
17 & 4.13 & 6.38$\pm$0.04 & 1.02$\pm$0.06 \\
18 & 4.30 & 6.72$\pm$0.04 & 0.34$\pm$0.06 \\
19 & 4.43 & 6.98$\pm$0.05 & 0.26$\pm$0.06 \\
20 & 4.61 & 7.35$\pm$0.05 & 0.37$\pm$0.07 \\
21 & 4.86 & 7.83$\pm$0.05 & 0.48$\pm$0.07 \\
22 & 5.06 & 8.23$\pm$0.05 & 0.40$\pm$0.07 \\
23 & 5.27 & 8.64$\pm$0.05 & 0.42$\pm$0.08 \\
24 & 5.60 & 9.30$\pm$0.06 & 0.66$\pm$0.08 \\
25 & 5.77 & 9.64$\pm$0.06 & 0.34$\pm$0.08 \\
\hline
\multicolumn{4}{c}{NGC 6946 -- S1 (IR)} \\
 2 & 1.69 & 1.53$\pm$0.03 & 1.23$\pm$0.05 \\
 3 & 1.83 & 1.81$\pm$0.03 & 0.28$\pm$0.05 \\
 4 & 2.22 & 2.59$\pm$0.03 & 0.77$\pm$0.05 \\
 5 & 2.53 & 3.21$\pm$0.03 & 0.62$\pm$0.05 \\
 6 & 2.86 & 3.86$\pm$0.03 & 0.65$\pm$0.05 \\
 7 & 3.13 & 4.41$\pm$0.03 & 0.55$\pm$0.05 \\
 8 & 3.30 & 4.73$\pm$0.03 & 0.32$\pm$0.05 \\
 9 & 3.44 & 5.01$\pm$0.03 & 0.27$\pm$0.05 \\
10 & 3.62 & 5.36$\pm$0.04 & 0.36$\pm$0.05 \\
11 & 3.80 & 5.74$\pm$0.04 & 0.38$\pm$0.05 \\
12 & 4.04 & 6.21$\pm$0.04 & 0.47$\pm$0.06 \\
13 & 4.25 & 6.63$\pm$0.04 & 0.42$\pm$0.06 \\
14 & 4.86 & 7.82$\pm$0.05 & 1.19$\pm$0.07 \\
15 & 5.43 & 8.97$\pm$0.05 & 1.14$\pm$0.08 \\
16 & 5.54 & 9.19$\pm$0.06 & 0.22$\pm$0.08 \\
17 & 5.89 & 9.88$\pm$0.06 & 0.69$\pm$0.08 \\
\hline
\multicolumn{4}{c}{NGC 6946 -- S2} \\
 2 & 1.27 & 0.70$\pm$0.02 & 0.27$\pm$0.02 \\
 3 & 1.48 & 1.11$\pm$0.02 & 0.41$\pm$0.02 \\
 4 & 1.74 & 1.63$\pm$0.02 & 0.52$\pm$0.03 \\
 5 & 2.09 & 2.33$\pm$0.02 & 0.69$\pm$0.03 \\
 6 & 2.41 & 2.96$\pm$0.02 & 0.63$\pm$0.03 \\
\hline 
\end{tabular}
\end{table}

\setcounter{table}{0}
\begin{table}
\caption{Continued}
\centering
\begin{tabular}{rrrr}
\hline \hline
$n$ & $r_n$ (kpc) & \multicolumn{1}{c}{$s_n$ (kpc)} &
\multicolumn{1}{c}{$l_n$ (kpc)} \\
\hline
\multicolumn{4}{c}{NGC 6946 -- S2} \\
 7 & 2.58 & 3.31$\pm$0.03 & 0.35$\pm$0.04 \\
 8 & 2.92 & 3.98$\pm$0.03 & 0.67$\pm$0.04 \\
 9 & 3.13 & 4.41$\pm$0.03 & 0.43$\pm$0.04 \\
10 & 3.30 & 4.73$\pm$0.03 & 0.32$\pm$0.05 \\
11 & 3.51 & 5.15$\pm$0.07 & 0.41$\pm$0.10 \\
12 & 3.73 & 5.58$\pm$0.04 & 0.44$\pm$0.05 \\
13 & 3.88 & 5.89$\pm$0.04 & 0.31$\pm$0.06 \\
14 & 4.04 & 6.21$\pm$0.04 & 0.32$\pm$0.06 \\
15 & 4.25 & 6.63$\pm$0.04 & 0.42$\pm$0.06 \\
16 & 4.43 & 6.98$\pm$0.09 & 0.35$\pm$0.13 \\
17 & 4.95 & 8.02$\pm$0.05 & 1.04$\pm$0.07 \\
18 & 5.11 & 8.33$\pm$0.05 & 0.30$\pm$0.07 \\
19 & 5.32 & 8.75$\pm$0.05 & 0.42$\pm$0.08 \\
20 & 5.66 & 9.41$\pm$0.06 & 0.66$\pm$0.08 \\
21 & 5.89 & 9.88$\pm$0.06 & 0.47$\pm$0.08 \\
\hline
\multicolumn{4}{c}{NGC 6946 -- S2 (IR)} \\
 2 & 1.24 & 0.65$\pm$0.03 & 0.35$\pm$0.05 \\
 3 & 1.69 & 1.53$\pm$0.03 & 0.88$\pm$0.05 \\
 4 & 1.98 & 2.12$\pm$0.03 & 0.59$\pm$0.05 \\
 5 & 2.13 & 2.41$\pm$0.03 & 0.29$\pm$0.05 \\
 6 & 2.58 & 3.31$\pm$0.03 & 0.90$\pm$0.05 \\
 7 & 2.72 & 3.58$\pm$0.03 & 0.27$\pm$0.05 \\
 8 & 2.98 & 4.10$\pm$0.03 & 0.52$\pm$0.05 \\
 9 & 3.30 & 4.73$\pm$0.03 & 0.63$\pm$0.05 \\
10 & 3.61 & 5.36$\pm$0.04 & 0.63$\pm$0.05 \\
11 & 4.08 & 6.29$\pm$0.04 & 0.93$\pm$0.06 \\
12 & 4.25 & 6.63$\pm$0.04 & 0.34$\pm$0.06 \\
13 & 4.48 & 7.07$\pm$0.04 & 0.44$\pm$0.06 \\
14 & 4.90 & 7.92$\pm$0.05 & 0.85$\pm$0.07 \\
15 & 5.16 & 8.43$\pm$0.05 & 0.51$\pm$0.07 \\
16 & 5.60 & 9.30$\pm$0.06 & 0.87$\pm$0.08 \\
\hline
\multicolumn{4}{c}{NGC 6946 -- S3} \\
 2 & 1.78 & 0.65$\pm$0.02 & 0.34$\pm$0.03 \\
 3 & 1.91 & 0.92$\pm$0.02 & 0.26$\pm$0.03 \\
 4 & 2.04 & 1.16$\pm$0.02 & 0.24$\pm$0.03 \\
 5 & 2.13 & 1.33$\pm$0.02 & 0.17$\pm$0.03 \\
 6 & 2.34 & 1.74$\pm$0.02 & 0.41$\pm$0.03 \\
 7 & 2.49 & 2.03$\pm$0.02 & 0.29$\pm$0.04 \\
 8 & 2.65 & 2.35$\pm$0.03 & 0.31$\pm$0.04 \\
 9 & 2.94 & 2.91$\pm$0.03 & 0.57$\pm$0.04 \\
10 & 3.24 & 3.48$\pm$0.03 & 0.57$\pm$0.04 \\
11 & 3.63 & 4.25$\pm$0.04 & 0.77$\pm$0.05 \\
12 & 3.83 & 4.63$\pm$0.08 & 0.38$\pm$0.11 \\
13 & 4.03 & 5.03$\pm$0.04 & 0.40$\pm$0.06 \\
14 & 4.38 & 5.71$\pm$0.04 & 0.68$\pm$0.06 \\
15 & 4.62 & 6.17$\pm$0.05 & 0.46$\pm$0.07 \\
16 & 4.82 & 6.56$\pm$0.05 & 0.38$\pm$0.07 \\
17 & 4.97 & 6.85$\pm$0.05 & 0.30$\pm$0.07 \\
18 & 5.24 & 7.37$\pm$0.05 & 0.52$\pm$0.07 \\
19 & 5.58 & 8.04$\pm$0.06 & 0.66$\pm$0.08 \\
\hline
\multicolumn{4}{c}{NGC 6946 -- S3 (IR)} \\
 2 & 1.74 & 0.58$\pm$0.03 & 0.43$\pm$0.05 \\
 3 & 1.96 & 1.00$\pm$0.03 & 0.41$\pm$0.05 \\
 4 & 2.19 & 1.46$\pm$0.03 & 0.46$\pm$0.05 \\
 5 & 2.39 & 1.83$\pm$0.03 & 0.37$\pm$0.05 \\
 6 & 2.65 & 2.35$\pm$0.03 & 0.51$\pm$0.05 \\
 7 & 2.91 & 2.85$\pm$0.03 & 0.51$\pm$0.05 \\
 8 & 3.17 & 3.35$\pm$0.03 & 0.50$\pm$0.05 \\
 9 & 3.45 & 3.89$\pm$0.03 & 0.54$\pm$0.05 \\
10 & 3.79 & 4.55$\pm$0.04 & 0.66$\pm$0.05 \\
\hline
\end{tabular}
\end{table}

\setcounter{table}{0}
\begin{table}
\caption{Continued}
\centering
\begin{tabular}{rrrr}
\hline \hline
$n$ & $r_n$ (kpc) & \multicolumn{1}{c}{$s_n$ (kpc)} &
\multicolumn{1}{c}{$l_n$ (kpc)} \\
\hline
\multicolumn{4}{c}{NGC 6946 -- S3 (IR)} \\
11 & 3.99 & 4.95$\pm$0.04 & 0.40$\pm$0.06 \\
12 & 4.25 & 5.45$\pm$0.04 & 0.50$\pm$0.06 \\
13 & 4.43 & 5.80$\pm$0.04 & 0.35$\pm$0.06 \\
14 & 4.92 & 6.75$\pm$0.05 & 0.95$\pm$0.07 \\
15 & 5.35 & 7.59$\pm$0.05 & 0.84$\pm$0.08 \\
16 & 5.52 & 7.92$\pm$0.06 & 0.33$\pm$0.08 \\
17 & 5.82 & 8.50$\pm$0.06 & 0.58$\pm$0.08 \\
\hline
\multicolumn{4}{c}{NGC 6946 -- S4} \\
 2 & 1.86 &  0.80$\pm$0.02 & 0.59$\pm$0.03 \\
 3 & 2.06 &  1.20$\pm$0.02 & 0.40$\pm$0.03 \\
 4 & 2.41 &  1.88$\pm$0.02 & 0.68$\pm$0.03 \\
 5 & 2.60 &  2.24$\pm$0.03 & 0.36$\pm$0.04 \\
 6 & 2.88 &  2.80$\pm$0.03 & 0.56$\pm$0.04 \\
 7 & 3.20 &  3.41$\pm$0.03 & 0.62$\pm$0.05 \\
 8 & 3.52 &  4.03$\pm$0.07 & 0.62$\pm$0.10 \\
 9 & 3.87 &  4.71$\pm$0.04 & 0.68$\pm$0.06 \\
10 & 4.03 &  5.03$\pm$0.04 & 0.32$\pm$0.06 \\
11 & 4.25 &  5.45$\pm$0.04 & 0.42$\pm$0.06 \\
12 & 4.43 &  5.80$\pm$0.09 & 0.35$\pm$0.13 \\
13 & 4.87 &  6.65$\pm$0.05 & 0.85$\pm$0.07 \\
14 & 5.08 &  7.06$\pm$0.05 & 0.41$\pm$0.07 \\
15 & 5.47 &  7.81$\pm$0.06 & 0.75$\pm$0.08 \\
16 & 5.64 &  8.15$\pm$0.06 & 0.34$\pm$0.08 \\
17 & 6.01 &  8.86$\pm$0.06 & 0.71$\pm$0.09 \\
18 & 6.46 &  9.75$\pm$0.07 & 0.89$\pm$0.09 \\
19 & 6.88 & 10.56$\pm$0.07 & 0.81$\pm$0.10 \\
\hline
\multicolumn{4}{c}{NGC 6946 -- S4 (IR)} \\
 2 & 1.71 &  0.51$\pm$0.03 & 0.39$\pm$0.05 \\
 3 & 1.88 &  0.84$\pm$0.03 & 0.33$\pm$0.05 \\
 4 & 2.06 &  1.20$\pm$0.03 & 0.36$\pm$0.05 \\
 5 & 2.46 &  1.98$\pm$0.03 & 0.78$\pm$0.05 \\
 6 & 2.97 &  2.97$\pm$0.03 & 0.99$\pm$0.05 \\
 7 & 3.27 &  3.55$\pm$0.03 & 0.57$\pm$0.05 \\
 8 & 3.48 &  3.96$\pm$0.03 & 0.41$\pm$0.05 \\
 9 & 4.16 &  5.28$\pm$0.04 & 1.32$\pm$0.06 \\
10 & 4.38 &  5.71$\pm$0.04 & 0.43$\pm$0.06 \\
11 & 4.82 &  6.56$\pm$0.05 & 0.84$\pm$0.07 \\
12 & 5.03 &  6.96$\pm$0.05 & 0.40$\pm$0.07 \\
13 & 5.47 &  7.81$\pm$0.06 & 0.85$\pm$0.08 \\
14 & 5.82 &  8.50$\pm$0.06 & 0.69$\pm$0.08 \\
15 & 6.67 & 10.15$\pm$0.07 & 1.65$\pm$0.10 \\
16 & 6.88 & 10.56$\pm$0.07 & 0.41$\pm$0.10 \\
\hline
\end{tabular}
\end{table}

\end{document}